\documentclass[fleqn,usenatbib,useAMS]{mnras}

\usepackage{newtxtext,newtxmath}
\usepackage[T1]{fontenc}
\usepackage{url}
\usepackage{multirow}
\usepackage{booktabs}
\usepackage{bm}
\usepackage{amsmath}	% Advanced maths commands

\DeclareRobustCommand{\VAN}[3]{#2}
\let\VANthebibliography\thebibliography
\def\thebibliography{\DeclareRobustCommand{\VAN}[3]{##3}\VANthebibliography}

\usepackage{graphicx}	% Including figure files
\usepackage{amsmath}	% Advanced maths commands
\usepackage{amssymb}	% Extra maths symbols
\usepackage{multicol}        % Multi-column entries in tables
\usepackage{bm}		% Bold maths symbols, including upright Greek
\usepackage{pdflscape}	% Landscape pages
\usepackage{tablefootnote}
\usepackage{threeparttable}
\usepackage{footnote}
\title[]{The Carbon-to-H$_2$, CO-to-H$_2$ Conversion Factors and Carbon Abundance on Kiloparsec Scales in Nearby Galaxies}

\author[Q. Jiao et al.]{
Qian Jiao$^{1}$\thanks{E-mail:jiaoqian@pmo.ac.cn},
Yu Gao$^{2,1}$\thanks{E-mail:yugao@xmu.edu.cn},
Yinghe Zhao$^{3}$%,
\\
$^{1}$Purple Mountain Observatory \& Key Lab of Radio Astronomy, Chinese Academy of Sciences (CAS), Nanjing 210023, China\\
$^{2}$Department of Astronomy, Xiamen University, Xiamen, Fujian 361005, China\\
$^{3}$Yunnan Observatories \& Key Laboratory for the Structure and Evolution of Celestial Objects, CAS, Kunming 650011, China
}

\date{Accepted XXX. Received YYY; in original form ZZZ}

\pubyear{****}

\newcommand{\CI}{[C\,{\sc i}]}
\newcommand{\CII}{[C\,{\sc ii}]}
\newcommand{\Cone}{[C\,{\sc i}]\,(1$-$0)}
\newcommand{\Ctwo}{[C\,{\sc i}]\,(2$-$1)}
\newcommand{\COone}{CO\,(1$-$0)}
\newcommand{\alCO}{$\alpha_\mathrm{CO}$}
\newcommand{\alCone}{$\alpha_\mathrm{[CI](1-0)}$}
\newcommand{\alCtwo}{$\alpha_\mathrm{[CI](2-1)}$}

\newcommand{\mum}{$\mu$m}
\newcommand{\HI}{H{\sc i}}
\newcommand{\LCOone}{$L'_{\rm {CO\,(1-0)}}$}
\newcommand{\LCone}{$L'_{\rm {[C\,{\sc I}]\,(1-0)}}$}

\begin{document}
\label{firstpage}
\pagerange{\pageref{firstpage}--\pageref{lastpage}}
\maketitle

\begin{abstract}
Using the atomic carbon \CI\,($^{3} \rm P_{1} \rightarrow {\rm ^3 P}_{0}$) and \CI\,($^{3} \rm P_{2} \rightarrow {\rm ^3 P}_{1}$) emission (hereafter \Cone\ and \Ctwo, respectively) maps observed with the $Herschel\ Space\ Observatory$, and \COone, \HI,  infrared and submm maps from literatures, we estimate the \CI-to-H$_2$ and CO-to-H$_2$ conversion factors of \alCone, \alCtwo, and \alCO\ at a linear resolution $\sim1\,$kpc scale for six nearby galaxies of M~51, M~83, NGC~3627, NGC~4736, NGC~5055, and NGC~6946. This is perhaps the first effort, to our knowledge, in calibrating both \CI-to-H$_2$ conversion factors across the spiral disks at spatially resolved $\sim1\,$kpc scale though such studies have been discussed globally in galaxies near and far. In order to derive the conversion factors and achieve these calibrations, we adopt three different dust-to-gas ratio (DGR) assumptions which scale approximately with metallicity taken from precursory results. We find that for all DGR assumptions, the \alCone, \alCtwo, and \alCO\ are mostly flat with galactocentric radii, whereas both \alCtwo\ and \alCO\ show decrease in the inner regions of galaxies. And the central \alCO\ and \alCtwo\ values are on average $\sim 2.2$ and $1.8$ times lower than its galaxy averages. The obtained carbon abundances from different DGR assumptions show flat profiles with galactocentric radii, and the average carbon abundance of the galaxies is comparable to the usually adopted value of $3 \times 10^{-5}$. We find that both metallicity and infrared luminosity correlate moderately with the \alCO\ whereas only weakly with either the \alCone\ or carbon abundance, and not at all with the \alCtwo.
%We find both metallicity and infrared luminosity have good, weak and no correlations with \alCO, \alCone\ and \alCtwo, respectively. And carbon abundance has weak correlation with metallicity and infrared luminosity.

\end{abstract}


%%%%%%%%%%%%%%%%%%%%%%%%%%%%%%%%%%%%%%%%%%%%%%%%%%

%%%%%%%%%%%%%%%%% BODY OF PAPER %%%%%%%%%%%%%%%%%%

\section{Introduction}

Molecular hydrogen (H$_2$), as the most abundant molecule, is the main fuel of star formation and plays a central role in the evolution of galaxies \citep{Kennicutt Evans 2012}. However, H$_2$ has no permanent dipole moment and is difficult to observe in emission. Other molecular gas tracers, such as carbon monoxide (CO), and neutral atomic carbon (\CI) are commonly used to study the molecular interstellar medium (e.g., \citealt{Papadopoulos et al. 2004amodel, Bolatto et al. 2013} and references therein). 

CO and CO-to-H$_2$ conversion factor (\alCO) in molecular clouds of galaxies near and far have been widely studied in theoretical models (e.g., \citealt{Narayanan et al. 2012, Feldmann et al. 2012, Bisbas et al. 2015, Bisbas et al. 2017, Papadopoulos et al. 2018}) and observations (e.g., \citealt{Solomon Barrett 1991, Yong Scoville 1991, Leroy et al. 2011, Sandstrom et al. 2013, Hunt et al. 2015, Shi et al. 2015, Israel 2020}). The \alCO\ proves to be sensitive with metallicity (see \citealt{Shi et al. 2016} for a review) and gas density, and appears to drop in the galaxy center (see \citealt{Bolatto et al. 2013} for a review). \citet{Sandstrom et al. 2013} concluded that the $\alpha_\mathrm{CO}$ value in the central region of most galaxies shows a factor of $\sim$2 times lower than the galaxy mean on average, and some can be factors of 5$-$10 below \citep{Solomon et al. 1997, Downes Solomon 1998} the ``standard" Milky Way (MW) value. While the only weak correlation between \alCO\ and metallicity in the sample of \cite{Sandstrom et al. 2013} indicating that the decreasing of  \alCO\ in the center might not be primary driven by metallicity but by other interstellar medium (ISM) conditions, e.g., high gas temperatures, large velocity dispersions.

While atomic carbon \CI\,($^{3} \rm P_{1} \rightarrow {\rm ^3 P}_{0}$) (rest frequency: 492.161 GHz, hereafter \Cone) and \CI\,($^{3} \rm P_{2} \rightarrow {\rm ^3 P}_{1}$) (rest frequency: 809.344 GHz, hereafter \Ctwo) lines received few attentions as molecular gas tracers because \CI\ was pictured emanating only from a narrow \CII/\CI/CO transition zone according to traditional photodissociation region (PDR) models \citep{Tielens et al. 1985, Hollenbach et al. 1991, Hollenbach et al. 1999}, and \CI\ emissions are difficult to observe with the ground-base facilities because the atmospheric transmissions at these frequencies are poor. However, some recent observations show that both \CI\ emissions correlate well with CO emission \citep{Ikeda et al. 1999, Shimajiri et al. 2013, Israel et al. 2015, Krips et al. 2016, Israel 2020}, and even perform well in tracing molecular gas in local infrared (IR) luminous objects \citep{Papadopoulos et al. 2004amodel, Jiao et al. 2017, Jiao et al. 2019} and high-redshift sub-millimeter galaxies (SMGs, \citealt{Alaghband-Zadeh et al. 2013, Yang et al. 2017}), as well as  now more frequently observed in main sequence of star-forming galaxies (SFGs) at $z>1$ \citep{Popping et al. 2017, Valentino et al. 2018, Valentino et al. 2020, Boogaard et al. 2020}.
Theoretical models which include turbulent \citep{Offner et al. 2014, Glover et al. 2015}, metallicity \citep{Glover & Clark 2016}, and cosmic ray \citep{Bisbas et al. 2015, Bisbas et al. 2017, Papadopoulos et al. 2018, Gaches et al. 2019} also predict widespread \CI\ emission maps which are similar to CO maps, and \CI\ even appears to do a slightly better job in tracing the molecular structure at low extinctions, or high cosmic-ray environments where CO is severely depleted. Specifically, \CI\ lines can be powerful molecular gas tracers at high redshift \citep{Walter et al. 2011, Tomassetti et al. 2014, Bothwell et al. 2017, Emonts et al. 2018}.
  
However, the \CI-to-H$_2$ conversion factors have only been constrained in some nearby galaxies (\citealt{Israel 2020} and \citealt{Izumi et al. 2020} for galaxy centers; \citealt{Miyamoto et al. 2021} for the northern part of M~83), averaged globally over entire nearby galaxies and (ultra)-luminous infrared galaxies ((U)LIRGs; \citealt{Papadopoulos et al. 2004data, Jiao et al. 2017, Crocker et al. 2019}), and few high-redshift galaxies based on absorption lines of H$_2$ and \CI\ \citep{Heintz Watson 2020}. Meanwhile, some recent studies using emission lines, provided the carbon abundance estimates for high-redshift main sequence galaxies \citep{Valentino et al. 2018, Boogaard et al. 2020}. %Using samples of high-redshift with ranges of $z=1.9-3.4$ gamma-ray burst and quasar molecular gas absorbers, \citet{Heintz Watson 2020} found that \alCone\ scale linearly with metallicity: log \alCone\ $= -1.13\times {\rm log(Z/Z_{\sun})} + 1.33$. They further applied their \alCone\ function for a sample of emission-selected galaxies at $z \sim 0-5$, and found a remarkable agreement between the molecular gas masses inferred from the absorption-derived \alCone\ with the typical \alCO-based estimates. And thus they concluded that the absorption$-$derived \alCone\ can be used to probe the universal properties of molecular gas in the local and high-redshift universe.
Few theoretical models including turbulent \citep{Offner et al. 2014}, metallicity \citep{Glover & Clark 2016}, or cosmic ray \citep{Gaches et al. 2019} predict the \CI-to-H$_2$ conversion factors in clouds.
%\citet{Offner et al. 2014} and \citet{Gaches et al. 2019} derived the \CI-to-H$_2$ conversion factors with simulations including turbulent and cosmic rays, respectively. 
Most of these precursory \CI-to-H$_2$ conversion factor results are global characteristics averaged across whole galaxies since few spatially resolved observations are available in nearby galaxies. In our previous work (\citealt{Jiao et al. 2019}), we studied both \CI\ lines with a linear resolution around $\sim1\,$kpc for a sample of nearby galaxies observed with the $Herschel\ Space\ Observatory$ ($Herschel$; \citealt{Pilbratt et al. 2010}) Spectral and Photometric Imaging Receiver Fourier Transform Spectrometer(SPIRE/FTS; \citealt{Griffin et al. 2010, Swinyard et al. 2014}). We found almost linearly correlation between luminosity of $L'_\mathrm{CO(1-0)}$ with both $L'_\mathrm{[CI](1-0)}$ and $L'_\mathrm{[CI](2-1)}$,  and relatively constant distribution of $L'_\mathrm{[CI](1-0)}$/$L'_\mathrm{CO(1-0)}$ within galaxies. Considering that the $\alpha_\mathrm{CO}$ drops in the center, the \CI-to-H$_2$ conversion factors might also vary within a galaxy. Here we will further use the previous sample to calibrate the \CI-to-H$_2$ conversion factors among galaxies, which might be the first effort in calibrating both \CI-to-H$_2$ conversion factors across the spiral disks at spatially resolved $\sim1\,$kpc scale though such studies have been discussed globally in galaxies near and far, and investigate the dependence of \CI-to-H$_2$ conversion factors on different physical properties of galaxies. %To our knowledge, this is perhaps the first efforts in calibrating the \CI-to-H$_2$ conversion factors across the spiral disks at spatially resolved $\sim1\,$kpc scale though such studies have been discussed globally in galaxies near and far.

%%Using samples of high-redshift gamma-ray burst and quasar molecular gas absorbers with ranges of $z=1.9-3.4$, \citet{Heintz Watson 2020} found that \alCone\ scales linearly with metallicity as: log$\,$\alCone\ $= -1.13\times {\rm log(Z/Z_{\sun})} + 1.33$. They further applied their \alCone\ function for a sample of emission-selected galaxies at $z \sim 0-5$, and found a remarkable agreement between the molecular gas masses inferred from their absorption-derived \alCone\ with the typical \alCO-based estimations. And thus they concluded that the absorption$-$derived \alCone\ can be used to probe the universal properties of molecular gas in the local and high-redshift universe. The simulation in \citet{Glover & Clark 2016} also demonstrated that the \alCone\ scales approximately with metallicity as \alCone\ $ \propto Z^{-1}$ in star-forming clouds. %We have presented the relation of log$\,$\alCone\ $= -1.13\times {\rm log(Z/Z_{\sun})} + 1.33$ as black line in the right-middle panel of Figure\,\ref{fig:acociwithparas}. Our derived \alCone\ distributes around their best-fit linear relation. 

This paper is organized as follows. Section 2 describes the solution methods of calibrating \CI-to-H$_2$ and CO-to-H$_2$ conversion factors. We give the details about the sample and data reduction in section 3. In section 4, we present the results and analysis. In the last section we summarize the main conclusions. %Throughout the paper, we use a Hubble constant of $H_0 = 70\mathrm{\ km\ s^{-1}\ Mpc^{-1}}$, $\Omega_\mathrm{M}=0.3$ and $\Omega_\mathrm{\lambda}=0.7$.

\section{Solution methods}

Assuming that dust and gas are well mixed and linearly related by dust-to-gas ratio (DGR) with the following equation: 
\begin{equation}
\frac{M_{\rm dust}}{{\rm DGR}} = M_{\rm gas} = 1.36 \times (M_{\rm HI} + M_{\rm H_2}),
\label{equ_dustmass_equation}
\end{equation} 
where $M_{\rm dust}$ is the dust mass which can be obtained with infrared data, and $M_{\rm HI}$ and $M_{\rm H_2}$ are the masses of atomic and molecular gas along the line of sight. The factor 1.36 accounts for helium and heavier elements. 

Using CO-to-${\rm H_2}$ conversion factor of \alCO, the molecular gas mass can be written as: $M_{\rm H_2} = \alpha_{\rm CO} L'_{\rm CO}$. 
Similarly the molecular gas mass can be obtained using \CI\ lines with:
	$M_{\rm H_2} = \alpha_{\rm [CI](1-0)} L'_{\rm [CI](1-0)}=\alpha_{\rm [CI](2-1)} L'_{\rm [CI](2-1)}$, where \alCone\ and \alCtwo\ are \CI-to-${\rm H_2}$ conversion factors.
Substituting $M_{\rm H_2}$, equation\,\ref{equ_dustmass_equation} becomes:
\begin{eqnarray}
\frac{M_{\rm dust}}{{\rm 1.36 \times DGR}} &=&M_{\rm HI} + M_{\rm H_2}\nonumber \\
&=&M_{\rm HI} + \alpha_{\rm CO} L'_{\rm CO}\nonumber \\
&=&M_{\rm HI} + \alpha_{\rm [CI](1-0)} L'_{\rm [CI](1-0)}\nonumber \\
&=&M_{\rm HI} + \alpha_{\rm [CI](2-1)} L'_{\rm [CI](2-1)}.
\label{equ_estimate_method}
\end{eqnarray}
Once $M_{\rm {dust}}$ is derived, we can then estimate \alCO\ and \CI\ conversion factors of $\alpha_{\rm [CI]}$ with DGR after assembling $M_{\rm {HI}}$, \LCOone, and \CI\ luminosities of $L'_{\rm [CI]}$.

Equations\,\ref{equ_dustmass_equation},\,\ref{equ_estimate_method} are mainly used globally for galaxies with some spatially resolved applications in CO. \citet{Leroy et al. 2011} estimated \alCO\ and DGR simultaneously for five local group galaxies by assuming that the DGR should be constant over a region of a galaxy. They divided each galaxy into several zones, and solved for \alCO\ which allows a single DGR to best describe each zone with maps of CO, \HI, and IR. Using a similar technique developed by \citet{Leroy et al. 2011}, \citet{Sandstrom et al. 2013} solved for \alCO\ and DGR simultaneously for 26 nearby star-forming galaxies with the assumption that the DGR is approximately constant on kiloparsec scales. They used high-resolution IR, CO and \HI\ maps to divide each galaxy into $\sim$kpc regions which they called $``$solution pixels$"$. And for each sampling point $i$ in their $``$solution pixels$"$, they derived the DGR$_i$ with an assumed \alCO. They then adjusted \alCO\ until they found the value that returns the most uniform DGR for all the DGR$_i$ values in the $``$solution pixels$"$. They finally found that the DGRs are well-correlated with metallicity with an approximately linear slope.

Based on Equations\,\ref{equ_dustmass_equation},\,\ref{equ_estimate_method}, here we mainly aim to solve for the \CI\ conversion factors by using the very limited spatially resolved \CI\ mapping from \citet{Jiao et al. 2019} observed with the $Herschel$ SPIRE/FTS, and estimate the CO conversion factor at the same time on the same scales. The SPIRE/FTS beams can be approximated as Gaussian profiles with FWHMs of 38.6$"$ and 36.2$"$ at 492 GHz and 809 GHz \citep{Makiwa et al. 2013}, respectively. With the low resolution of \CI\ observations, it is impossible to divid a galaxy into properly $``$solution pixels$"$ as \citet{Sandstrom et al. 2013} and \citet{Leroy et al. 2011}. And thus it becomes necessary to assume a DGR to obtain the \CI\ or CO conversion factors. The adopted DGR assumptions are shown in section\,3.5.

\section{Sample and data reduction}

\begin{table*}
  \caption{The basic information of the sample galaxies}
  \label{tab:sample}
  \begin{tabular}{lcccccccccccc}
      \hline
      Name & R.A. & Dec. & D & T$^a$ &  Inclination$^b$ & P.A.$^b$ & $R_{25}$$^c$ & $B_{maj}$$^d$ & $B_{min}$$^d$ & BPA$^e$ & \CI\ Spatial Scale \\ 
      		 & J2000 & J2000 & (Mpc) & (Type) & ($\degr$) & ($\degr$)& ($'$) & ($''$) & ($''$) & ($\degr$) & (kpc) \\
      \hline
      M~51 & 13$:$29$:$52.7 & +47$:$11$:$42.6  & 8.2 & 4 &  42 & 172 & 5.61 & 11.92 & 10.01 & -86.0 & 1.5 \\
      M~83 & 13$:$37$:$00.9 & $-$29$:$51$:$55.5 & 4.7  & 5 & 24 & 225 & 6.74  & 15.16 & 11.44 & -3.0  & 0.9 \\
      NGC~3627 & 11$:$20$:$15.0 & +12$:$59$:$29.5 & 9.4 & 3 & 62 & 173 & 4.56 & 10.60 & 8.85 & -48.0 & 1.8 \\
      NGC~4736 & 12$:$50$:$53.1 & +41$:$07$:$13.7  & 4.7 & 2 & 41  & 296 & 5.61 & 10.22 & 9.07 & -23.0  & 0.9 \\
      NGC~5055 & 13$:$15$:$49.3 & +42$:$01$:$45.4 & 7.9 & 4 & 59 & 102 & 6.30 & 10.06 & 8.66 & -40.0 & 1.5 \\
      NGC~6946 & 20$:$34$:$52.3 & +60$:$09$:$14.1 & 6.8 & 6 & 33 & 243  & 5.74 & 6.04 & 5.61 & 6.6 & 1.3 \\
%      NGC~4826 & 12$:$56$:$43.6 & +21$:$40$:$58.7 & 7.5 & 2 & 65 & 121 & 5.00 & 12.18 & 9.35 & −76.1  \\
      \hline 
\end{tabular}
 \begin{tablenotes}
 	\footnotesize
\item $^a$  Morphological type T as given in the RC3 catalog from \citet{de Vaucouleurs et al. 1991}.\\
\item $^b$ The inclinations and position angles (P.A.) are adopted from \citet{Walter et al. 2008}. \\
\item $^c$ $R_{25}$ is the B-band isophotal radius at 25 mag arcsec$^{-2}$  adopting from \citet{Moustakas et al. 2010} except for M~83 which is adopted from \citet{Bresolin Kennicutt 2002}.\\
\item $^d$ Major and minor axis of synthesized beam of VLA in arcsec from \citet{Walter et al. 2008}.  \\
\item $^e$ Position angle of synthesized beam of VLA in degrees from \citet{Walter et al. 2008}.
 \end{tablenotes}  
\end{table*}

\subsection{Sample selection}
The galaxies are mainly selected from the cross matching of the sample in \citet{Jiao et al. 2019} which has available \CI\ and \COone\ maps with $Herschel$ SPIRE/FTS and Nobeyama 45-m telescope respectively, and the survey from ``The \HI\ Nearby Galaxies Survey" (THINGS; \citealt{Walter et al. 2008}). We primarily obtain eight galaxies. While for galaxies with high inclination, the observation of \HI\ will include contributions from gas at larger radii which have a lower DGR and be less molecular-gas-rich \citep{Leroy et al. 2011, Sandstrom et al. 2013}, and may not suit for this method. We thus excluding two galaxies (NGC~3521 and NGC~7331, inclinations are 73$\degr$ and 76$\degr$) with inclination lager than 65$\degr$ (similar to \citealt{Sandstrom et al. 2013}), and finally obtain six galaxies as shown in Table\ref{tab:sample}. For the adopted six galaxies, we further make use of the dust observations from LVL (``the Spitzer Local Volume Legacy"; \citealt{Dale et al. 2009}) survey. 

\subsection{\CI\ and \COone\ luminosities}

The details of the data reduction and luminosity estimation of \CI\ and \COone\ lines are shown in \citet{Jiao et al. 2019}. Briefly, the \CI\ lines are reduced by the standard SPIRE/FTS reduction and calibration pipeline for mapping mode included in the Herschel Interactive Processing Environment (HIPE; \citealt{Ott 2010}) version 14.1, and its fluxes are estimated by fitting the observed line profiles with the instrumental Sinc function (see details in \citealt{Lu et al. 2017}). For the positions without significant detections (signal-to-noise ratios of SNRs $< 3\sigma$), we estimate 3$\sigma$ as upper limits of line integrated intensities. The \COone\ lines are collected from \citet{Kuno et al. 2007}, and then smoothed and regridded to the same resolution (FWHMs of 38.6$''$) and pixel scale as \Cone\ with uncertainly of $\sim 36\%$. In Figure\,\ref{fig:CI_All}, we present the \CI\ integrated intensity distributions with contours of convolved \COone\ emission, and the minus values are 3$\sigma$ for the non-detections. The \Cone\ detection region is smaller than that of \Ctwo\ due to the reduced sensitivity near the low-frequency end of the SPIRE Long Wavelength Spectrometer Array \citep{Swinyard et al. 2014}. The luminosities of \CI\ and \COone\ are estimated using \citet{Papadopoulos et al. 2012a}:
\begin{equation}
	L'_x =  3.25 \times 10^3\,
	           \left [
	              \frac{D_\mathrm{L}^2 (\mathrm{Mpc})} {1+z}
	           \right ]
	           \left (
	               \frac{v_{x,\mathrm{rest}}} {100 GHz}
	           \right )^{-2}
	           \left [
	                \frac{\int_{\Delta \mathrm{V}} \, S_v \, d\mathrm{V}} 
	                       {Jy\, km\, s^{-1}}
	           \right ], 
\label{equ_luminosity}
\end{equation}
where $L'_x$ is in unit of $\mathrm{K\, km\, s^{-1}}$, $v_{x,\mathrm{rest}}$ is the rest frequency and $S_v$ represents the line flux density.

\subsection{\HI\ maps}

The \HI\ maps are obtained from \citet{Walter et al. 2008} observed with the Very Large Array (VLA) of the National Radio Astronomy Observatory, and then converted to \HI\ masses using Equation\,(3) of \citet{Walter et al. 2008}. Based on its individual FWHM beam properties as shown in Table\ref{tab:sample}, we produce circular Gaussian point spread functions (PSF) for each galaxies. And then we convolve the \HI\ images with convolution kernels generated by comparing the PSF profiles of each galaxy with SPIRE/FTS Gaussian profile of FWHM of 38.6$''$ \citep{Aniano et al. 2011}. We adopted uncertainty of 10$\%$ for \HI\ masses \citep{Walter et al. 2008}.

\subsection{Dust mass maps}

The dust mass maps are derived by fitting spectral energy distribution (SED) with the observation of $Spitzer$ and $Herschel$ IR and submm maps. The resolution of the obtained dust map is equivalent to the lowest resolution of the IR and submm maps. In order to compare the dust mass with \CI\ maps, the lowest resolution of the IR and submm maps should be better than that of \Cone\ (FWHM$\sim$38.6$''$). With the limited resolution, we finally adopt IRAC 3.6, 4.5, 5.8, and 8.0\,\mum; MIPS 24 and 70\,\mum; and $Herschel$ PACS 70, 100, and 160\,\mum; and SPIRE 250 and 350\,\mum\ data. Specially, the MIPS 24\,\mum\ of galaxy M\,83 has few saturated pixels in the central region. For IRAC data, we use aperture correction factors of 0.91, 0.94, 0.66, and 0.74 for the 3.6\,\mum, 4.5\,\mum, 5.8\,\mum, and 8.0\,\mum\ bands, respectively (IRAC Instrument Handbook Version 2.12\footnote{\url{https://irsa.ipac.caltech.edu/data/SPITZER/docs/irac/iracinstrumenthandbook/IRAC_Instrument_Handbook.pdf}}). The calibration uncertainties of IRAC are $5\%-10\%$ for 3.6 and 4.5\,\mum, and $10\%-15\%$ for 5.8 and 8.0\,\mum\ \citep{Reach et al. 2005, Farihi et al. 2008}, and $10\%$ IRAC uncertainties are adopted here. MIPS calibration uncertainties are $4\%$ and $5\%$ at 24 and 70\,\mum\ \citep{Engelbracht et al. 2007, Gordon et al. 2007, Stansberry et al. 2007}, respectively. For $Herschel$, the absolute calibration accuracies are $5\%$ and $7\%$ for PACS and SPIRE data (SPIRE Observers' Manual HERSCHEL-DOC-0798, version 2.4\footnote{\url{http://herschel.esac.esa.int/Docs/SPIRE/pdf/spire_om_v24.pdf}}), respectively.

The $Spitzer$ and $Herschel$ IR and submm maps are smoothed \citep{Aniano et al. 2011} and regridded to the same resolution and pixel scale of \Cone. We use a standard dust model developed by \citet{Draine Li 2007} with a Milky Way grain size distribution to estimate the dust mass maps. The \citet{Draine Li 2007} model describes the interstellar dust as a mixture of carbonaceous grains and amorphous silicate grains with following parameters: the dust mass $M_\mathrm{dust}$; the fraction of dust mass ($q_{\rm{PAH}}$) in the form of polycyclic aromatic hydrocarbon (PAH) grains with fewer than $10^3$ carbon atoms; the minimum ($U_{\rm{min}}$) intensity of the radiation field that responds to heating majority ($1-\gamma$) of the dust; and the other small fraction ($\gamma$) of dust exposed to starlight with power-law distribution of starlight intensities ranging from $U_{\rm{min}}$ to $U_{\rm{max}}$ which associate with PDRs; exponent ($\alpha$) of the power-law distribution of intensities between $U_{\rm{min}}$ to $U_{\rm{max}}$. Following \citet{Draine et al. 2007}, the dust mass $d M_\mathrm{dust}$ exposed to radiation intensities in [$U,\ U + dU$] can be expressed as a combination of a Dirac $\delta-$function and a power law:
\begin{eqnarray}
\frac{dM_\mathrm{dust}} {dU} &=& ( 1 - \gamma) M_\mathrm{dust} \delta(U-U_\mathrm{min}) \nonumber \\
	&+& \gamma	M_\mathrm{dust} \frac{(\alpha - 1)} 
	    {U_\mathrm{min}^{(1-\alpha)} - 
	     U_\mathrm{max}^{(1-\alpha)}} U^{-\alpha}. 
\label{equ_Drainemodel}
\end{eqnarray}

The dust model of distribution function has six adjustable parameters [$M_\mathrm{dust}$, $\gamma$, $U_\mathrm{min}$, $U_\mathrm{max}$, $\alpha$, $q_{\rm{PAH}}$]. We further add a 5000\,K black-body spectrum to represent the stellar emission which dominates at wavelengths smaller than 5 \mum\ \citep{Draine et al. 2007, Munoz-Mateos et al. 2009}. \citet{Draine et al. 2007} showed that $U_{\rm{max}}=10^6$ and $\alpha=2$ work well for a wide range of galaxies, and thus we also fix $U_{\rm{max}}=10^6$ and $\alpha=2$. We built a grid model with different $U_\mathrm{min}$, $q_{\rm{PAH}}$ and $\gamma$. The $U_\mathrm{min}$ is same as the original model of \citet{Draine et al. 2007} which ranges from 0.10 to 25.0. The $q_{\rm{PAH}}$ is linear interpolated in steps of 0.1\% with ranges from 0.47\% to 4.58\%, and $\gamma$ is from 0 to 1 in steps of 0.01. The allowed ranges for each parameter are shown in Table\,\ref{tab:parameters}.

Following \citet{Draine et al. 2007}, we add an additional $\sim10\%$ error at each band due to the limited accuracy of the model \citep{Draine et al. 2007, Munoz-Mateos et al. 2009}. We can therefore estimate $M_\mathrm{dust}$, $q_{\rm{PAH}}$, $\gamma$, and $U_{\rm{min}}$ with the $Spitzer$ and $Herschel$ IR and submm maps by finding the best-fit SED models. We look for the best fitting model by minimizing the reduced $\chi^2$. And the 1$\sigma$ uncertainty of each parameter is derived by projecting the overall $\chi^2$ distribution over the one-dimensional space of that parameter, and then looking for the values which satisfied $\chi^2 = {\chi_{\rm{min}}^2 +1}$ \citep{Press et al. 1992, Munoz-Mateos et al. 2009}. The obtained dust mass maps are shown in Figure\,\ref{fig:dustmass}. 
For galaxy M~83 which includes several saturations in the central region of the 24\,\mum\ image, we compare the dust mass derived with or without the 24\,\mum\ data. And we find that the dust mass changes little for these two methods. This might be due to that the new smoothed and regridded data reduce the influence of saturation. Meanwhile, the other ten bands used for the dust model fitting will wash out the influence of one band, and the saturation of 24\,\mum\ will be reflected in $\chi^2$. In the following analysis, we use the model result of M~83 which includes the band of MIPS 24\,\mum.

We also derive the average interstellar radiation field for each pixel by:
\begin{equation}
	\overline{U} = ( 1 - \gamma)U_\mathrm{min} + \gamma	U_\mathrm{min} \frac{\mathrm{ln}(U_\mathrm{max}/U_\mathrm{min})} {1-(U_\mathrm{min}/U_\mathrm{max})}, 
\label{equ_mean_intensity}
\end{equation}
with the best fitting model, and estimate the infrared luminosity $L_\mathrm{IR}$ of dust emission by integrating over the best fitting model from 8 to 1000 \mum. 

\begin{table}
  \caption{Allowed ranges for each parameter}
  \label{tab:parameters}
  \begin{tabular}{lccccc}
      \hline
      Parameter & Min & Max & Parameter Grid  \\ 
      \hline
      $q_\mathrm{PAH}$ &  0.47\% &  4.58\% & in steps $\Delta_\mathrm{PAH} =$ 0.1\% \\
      $\gamma$ & 0 & 1 & in steps $\Delta_\gamma = 0.01$ \\
      $U_\mathrm{min}$ & 0.10 & 25.0 & steps following DL07$^a$ \\
      $U_\mathrm{max}$ & $10^6$ & $10^6$ & fixed \\
      $\alpha$ & 2 & 2 & fixed \\
      $M_\mathrm{dust}$ & 0 & $\infty$ & continuous fit \\ 
      \hline
\multicolumn{4}{l}{$^a$ DL07 stands for \citet{Draine Li 2007}.}\\%, i.e., 0.10, 0.15, 0.20, 0.30, 0.40, 0.50, 0.70, 0.80, 1.00, 1.20, 1.50, 2.00, 2.50, 3.00, 4.00, 5.00, 7.00, 8.00, 10.0, 12.0, 15.0, 20.0, 25.0. \\
\end{tabular}
\end{table}

\subsection{Metallicity and DGR}

A correlation between DGR and the gas-phase oxygen abundance has been widely shown (e.g., \citealt{Issa et al. 1990, Lisenfeld Ferrara 1998, Edmunds 2001, James et al. 2002, Hirashita et al. 2002, Boissier et al. 2004, Draine Li 2007, Munoz-Mateos et al. 2009, Sandstrom et al. 2013, Remy-Ruyer et al. 2014, De Vis et al. 2017, De Vis et al. 2019, Peroux & Howk 2020}), and the DGR decreases with radius and following a trend with metallicity. \citet{Moustakas et al. 2010} derived the metallicity gradients for 21 SINGS (The $Spitzer$ Infrared Nearby Galaxies Survey, \citealt{Kennicutt et al. 2003}) galaxies with oxygen abundance computed using two different strong-line abundance calibrations: a theoretical (\citealt{KK04}, hereafter KK04) and an empirical \citep{PT05} calibrations. The values in KK04 tend to be higher than the values in \citet{PT05} by $\sim 0.6$ dex. Using the metallicity gradients derived from \citet{Moustakas et al. 2010} with oxygen abundance calibrated by KK04, \citet{Munoz-Mateos et al. 2009} found a linear correlation between the DGR and metallicity: 
\begin{equation}
	{\rm log(DGR) = 5.63 + 2.45 \times log(O/H)}. 
\label{equ_DGR_metallicity_MM09}
\end{equation}

With the assumption that the abundances of all heavy elements are proportional to the oxygen abundance and that all heavy elements condensed to form dust in the same way as in the MW, \citet{Draine Li 2007} scaled the dust-to-gas ratio proportionally to the oxygen abundance:
\begin{equation}
	{\rm \frac{M_{dust}}{M_{gas}} \approx \frac{0.01}{1.36} \frac{(O/H)}{(O/H)_{MW}}}. 
\label{equ_DGR_metallicity_DL07}
\end{equation}
where 0.01 is the DGR of the MW and the factor 1.36 accounts for helium and heavier elements. \citet{Munoz-Mateos et al. 2009} compared their derived DGR correlation with the correlation in \citet{Draine Li 2007} for their sample galaxies, and found that DGR in Sc-Sd spirals decreases faster than Sb-Sbc galaxies (see their Figure\,15). More specially, for Sb-Sbc galaxies the DGR values derived are more consistent with Equation\,\ref{equ_DGR_metallicity_DL07}, while for Sa-Sab and Sc-Sd galaxies Equation\,\ref{equ_DGR_metallicity_MM09} fit the derived values better (see the Figures\,15 and 16 in \citealt{Munoz-Mateos et al. 2009}).

\citet{Remy-Ruyer et al. 2014} and \citet{De Vis et al. 2019} are two representative galaxy-integrated studies for dust properties in the nearby universe. Using a sample of 126 galaxies over a 2 dex metallicity calibrated from \citet{PT05}, \citet{Remy-Ruyer et al. 2014} found a broken power law trend can best describe the gas-to-dust mass ratio as a function of metallicity with uncertain to a factor of 1.6. On the other hand, \citet{De Vis et al. 2019} found that a single power law provides the best description of DGR with global metallicity for a sample of $\sim 500$ galaxies, and they further estimated the power law fits for several widely used metallicity calibration (see Table\,4 in \citealt{De Vis et al. 2019}). However, the metallicities in both works correspond to global estimates, and range from $12 + {\rm log(O/H)} = 7.14$ to 9.10 with 30\% of the sample with $12 + {\rm log(O/H)} \leq 8.0$ in \citet{Remy-Ruyer et al. 2014}. The resolved metallicity  in our sample is calibrated with the same resolution as its \CI\ spatial scale, which is similar to the metallicity and dust scale of \citet{Munoz-Mateos et al. 2009}. Besides, our sample is overlap with the sample of \citet{Munoz-Mateos et al. 2009} except for M\,83. So in the following analysis, we mainly use the DGR calibration from \citet{Munoz-Mateos et al. 2009}.

We use metallicity gradients from \citet{Moustakas et al. 2010} with oxygen abundances in KK04 to estimate the DGR distribution of each galaxy. We adopt the radial gradient metallicity from Table\,8 of \citet{Moustakas et al. 2010} with KK04 calibration. For galaxy of NGC~3627 which has no available gradient measurements, we use fixed metallicity for the entire galaxy from Table\,9 of \citet{Moustakas et al. 2010}. Specifically, for galaxy M~83 which is not among the sample of \citet{Moustakas et al. 2010}, we adopt the gradient metallicity from \citet{Bresolin Kennicutt 2002} with no available errors. The adopted metallicities and gradients are shown in Table\,\ref{tab:metallicity}. In order to evaluate and analysis the conversion factors of \CI\ and \COone, we use three different assumptions of DGR estimation:

$(i)$ DGR from \citet{Munoz-Mateos et al. 2009}: We adopt Equation\,\ref{equ_DGR_metallicity_MM09} to derive the DGR for most of the samples, while for Sbc galaxies of M~51 and NGC~5055, the DGRs are estimated with Equation\,\ref{equ_DGR_metallicity_DL07}.

$(ii)$ DGR from \citet{Draine Li 2007}: We use Equation\,\ref{equ_DGR_metallicity_DL07} to estimate the DGR for each galaxy.

$(iii)$ DGR from \citet{Sandstrom et al. 2013}: We also consider the DGR derived from \citet{Sandstrom et al. 2013} for each galaxy: ${\rm log(DGR) = -1.86 + 0.87( 12 + log(O/H) - 9.05)}$.\\

\begin{table}
 % \centering
  \caption{The adopted metallicity of each galaxy}
  \label{tab:metallicity}
  \begin{threeparttable}
  \begin{tabular}{lcc}
      \hline
      Name & Central\_Z$^a$ & Gradient\_Z$^b$  \\ 
      		& (KK04) & (KK04) \\
      \hline
      M~51 & $9.33 \pm 0.01$ & $-0.50 \pm 0.05$\\
      M~83 & 9.07  & $-0.186$ \\
      NGC~3627 & $ 8.99 \pm 0.10$ & ...\\
      NGC~4736 & $9.04 \pm 0.01$ & $-0.11 \pm 0.15$\\
      NGC~5055 & $9.30\pm 0.04$ & $ -0.54 \pm 0.18 $\\
      NGC~6946 & $9.13 \pm 0.04$ & $-0.28 \pm 0.10$\\
      \hline
\end{tabular}
\begin{tablenotes}
\footnotesize
\item $^a$ Central oxygen abundance based on the derived abundance gradient with \citet{KK04} calibration except for M~83 which is adopted from \citet{Bresolin Kennicutt 2002}.\\
\item $^b$ Slope of the radial abundance gradient with \citet{KK04} calibration.	
\end{tablenotes}
    \end{threeparttable}

\end{table}

\section{Results and Analysis}

\subsection{Calibration of \CI\ and CO conversion factors}

\begin{figure*}
 \begin{center}
  \includegraphics[bb = 74 125 732 472, clip, width=0.33\textwidth]{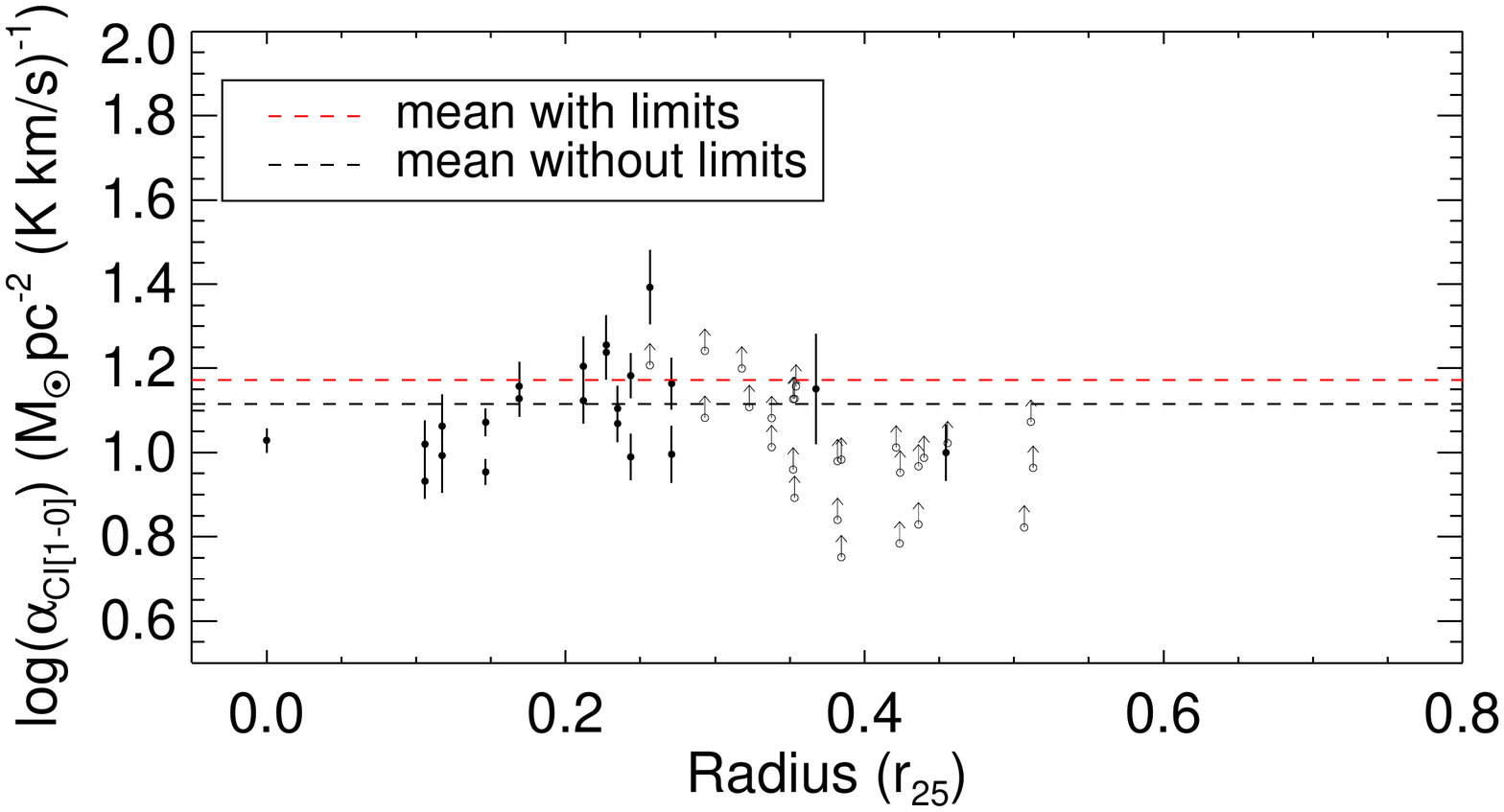} 
  \includegraphics[bb = 74 125 732 472, clip, width=0.33\textwidth]{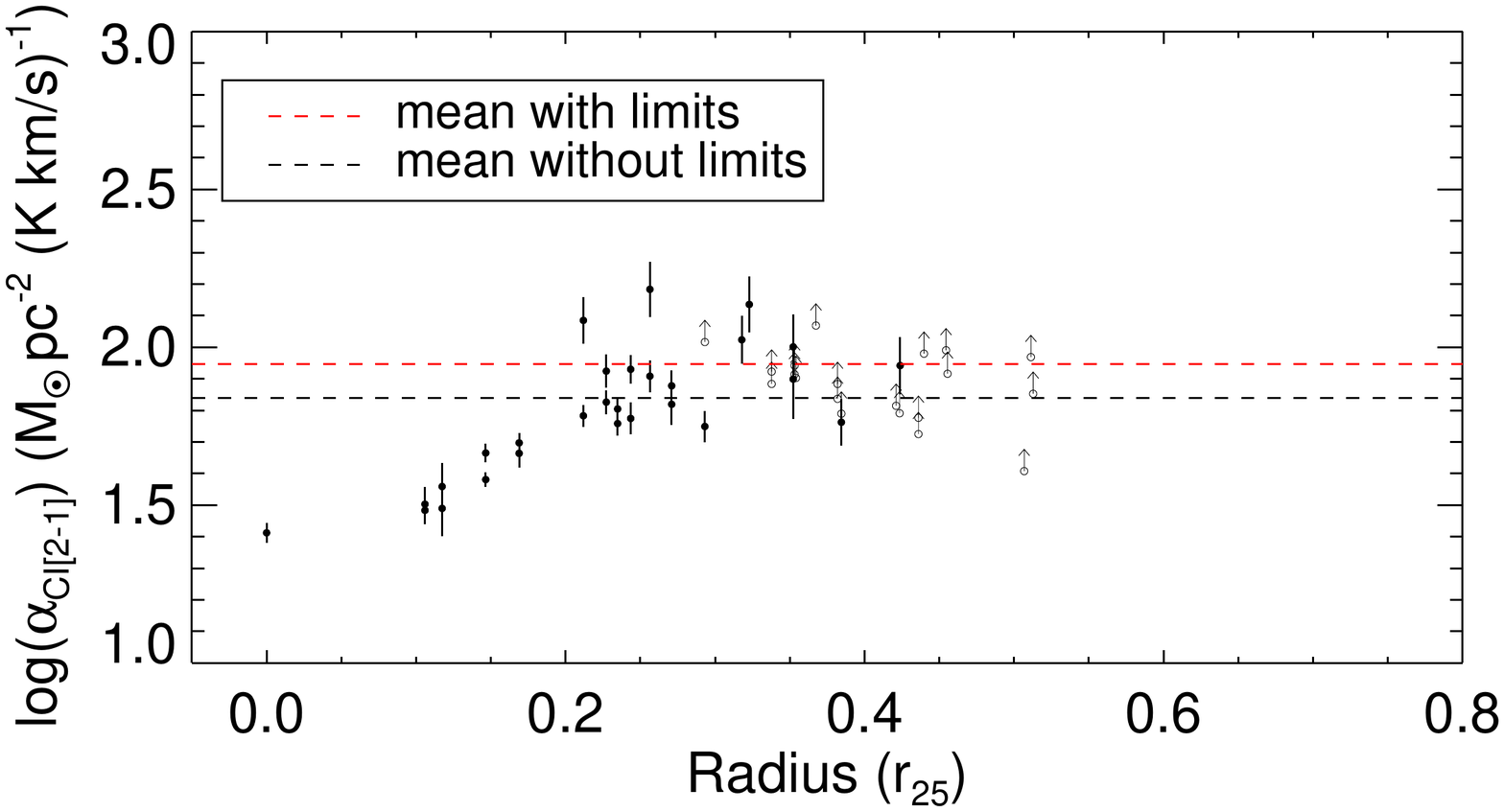} 
  \includegraphics[bb = 74 125 732 472, clip, width=0.33\textwidth]{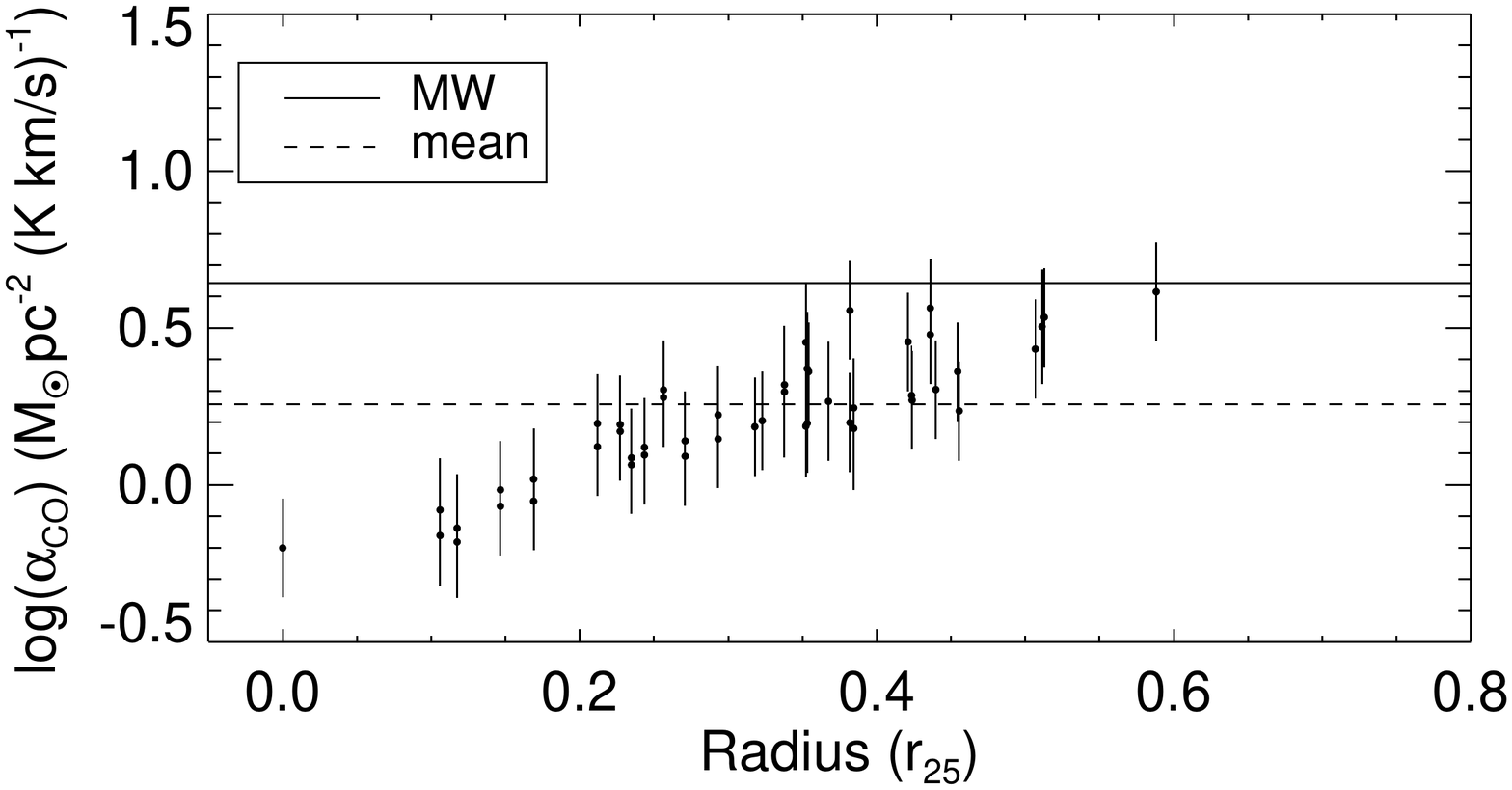} 
 \end{center}
\caption{\alCone\ (left panel), \alCtwo\ (middle panel), and \alCO\ (right panel) for NGC~6946 as functions of galactocentric radii ($r_{25}$). The filled black circles represent the detection points, and the black circles show the lower limits for the non-detections. The dotted black and red lines in each panels show the average values without and with lower limits respectively, and the solid line in the right panel shows the MW value of \alCO\ = 4.4\,${\rm M_{\sun} pc^{-2} (K\ km\ s^{-1})^{-1} }$.}\label{fig:aciwithradius}
\end{figure*}

Throughout the paper, we define $r_{25}= r / \rm{R_{25}}$, where $r$ is the galactocentric radius corrected with position and inclination angles listed in Table\ref{tab:sample}, and $\rm{R_{25}}$ is the B-band isophotal radius at 25 mag arcsec$^{-2}$ shown in Table\ref{tab:sample}. We then derive the distributions of \alCone, \alCtwo, and \alCO\ with Equation\,\ref{equ_estimate_method} using assumption of DGR$(i)$ for each galaxy. For the non-detections of \CI\ regions, we estimate the lower limits of $\alpha_{\rm [CI]}$ using $3\sigma$ of $L'_{\rm [CI]}$.
In Figure\,\ref{fig:aciwithradius}, we present the results of NGC~6946 as an example. The results for the other galaxies are shown in Figure\,\ref{fig:append_radii} in Appendix. Few outliers of \alCtwo\ with large $r_{25}$ especially for galaxy NGC~3627 might be due to that these pixels mainly locate around the boundary of SPIRE/FTS observations (see the \Ctwo\ maps in Figure\,\ref{fig:CI_All}) and have low SNRs ($\sim 3.2 - 3.4$) comparing with other points which can even reach SNRs $\sim65$.

From left to right panels, Figure\,\ref{fig:aciwithradius} shows \alCone, \alCtwo, and \alCO\ as a function of galactocentric radii in unit of $r_{25}$. The filled black circles represent the detection points, and the black circles show the lower limits for the non-detections. The dotted black lines in each panels show the average values for detections, and the solid line in the right panel shows the MW value of \alCO\ = 4.4\,${\rm M_{\sun} pc^{-2} (K\ km\ s^{-1})^{-1} }$ for comparison. We also estimate the average values of \alCone\ and \alCtwo\ that also take into consideration all the lower limits using $enparCensored$ function in the EnvStats package within the \rm{R}\footnote{http://www.R-project.org/} statistical software environment, and present as dotted red lines in Figure\,\ref{fig:aciwithradius}. There is a moderate trend for lower \alCO\ at smaller radii ($r_{25}<0.3$), and the central \alCO\ ($0.6\,{\rm M_{\sun} pc^{-2} (K\ km\ s^{-1})^{-1} }$) shows several times lower than the MW \alCO, which agrees well with \citet{Sandstrom et al. 2013}. The right panels of Figure\,\ref{fig:aciwithradius} and Figure\,\ref{fig:append_radii} only present \alCO\ in the region same as \CI\ observations. In Table\,\ref{tab:alcicovalues}, we list the central and average \alCO\ values of all CO detections for each galaxy.

The \alCtwo\ of NGC~6946 also shows a weak correlation with radius in the inner region, and becomes flat at larger radii. As presented in Table\,\ref{tab:alcicovalues}, the central \alCtwo\ tends to be almost three times lower than its galaxy average, and becomes even more times lower than its galaxy average value when the lower limits are considered. There is no obvious correlation between \alCone\ and radius, while the central \alCone\ is slightly smaller than the galaxy averages.

%NGC~3627.... no metallicity gradient, constant DGR gives the similar results for each other. The \alCO\ for each galaxies shows 

\subsection{Properties of \CI\ and CO conversion factors}

\begin{figure*}
 \begin{center}
  \includegraphics[bb = 67 115 730 490, width=0.95\textwidth]{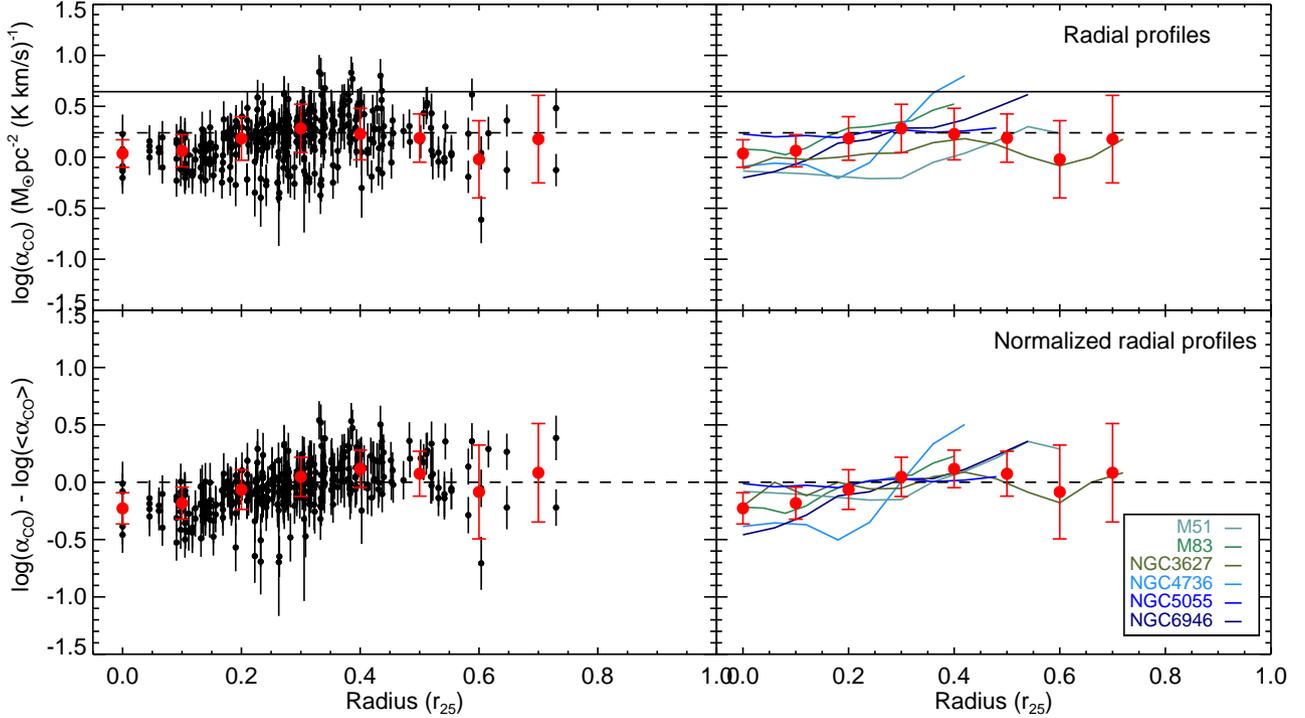} 
 \end{center}
\caption{The top left panel shows the \alCO\ as a function of galactocentric radii for all spaxels of our sample galaxies together, and the bottom left panel shows the same values normalized by each galaxy-averaged \alCO. The filled black circles represent each points for the whole sample. The radial profile and normalized profile of each galaxy are shown with colored lines in the top and bottom right panels, respectively. In each panels, the mean and standard deviation of all points in 0.1$r_{25}$ bin are shown by red symbols, and the dotted lines show the average values. The solid line in the top panels show the MW value of \alCO\ = 4.4\,${\rm M_{\sun} pc^{-2} (K\ km\ s^{-1})^{-1} }$.}\label{fig:acowithradius_together}
\end{figure*}

\begin{figure*}
 \begin{center}
  \includegraphics[bb = 67 115 730 490, width=0.95\textwidth]{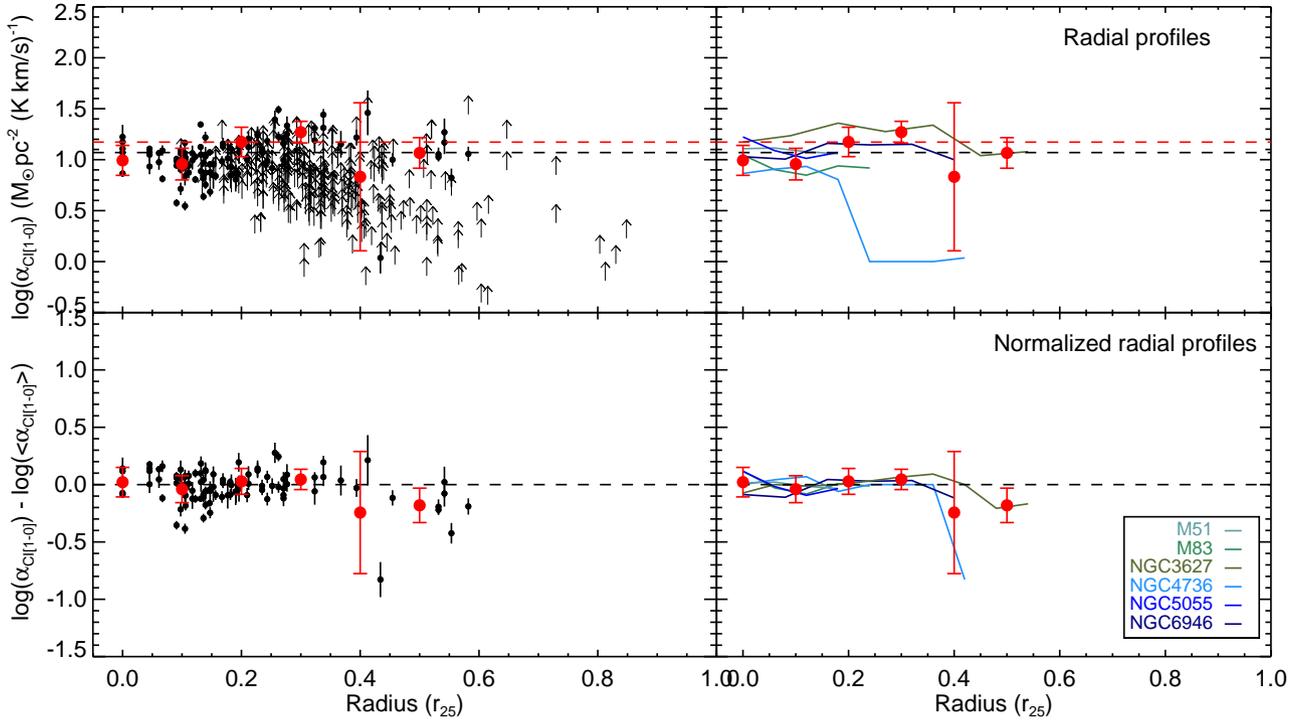} 
 \end{center}
\caption{The left panels show the \alCone\ for each galaxy together as a functions of galactocentric radii, and filled black circles represent each detections for the whole sample, and the arrows show the lower limits of \alCone. The right panels show the radial profile for the detections of each galaxies. The top panels show the original results, and the bottom panels show the same values normalized by its galaxy-averaged \alCone\ for detections. The dotted black lines show the average values of detections for galaxies together (\alCone\ $= 11.7 \pm 5.2 M_{\odot }\,{\rm pc^{-2}\,(K\,km\,s^{-1})^{-1}}$), and the dotted red lines show the average values with limits (\alCone\ $= 14.9 \pm 6.5 M_{\odot }\,{\rm pc^{-2}\,(K\,km\,s^{-1})^{-1}}$). The mean and standard deviation of all detections in 0.1$\,r_{25}$ bins are shown as red filled circles.}
\label{fig:aciwithradius_together}
\end{figure*}

\begin{figure*}
 \begin{center}
  \includegraphics[bb = 67 115 730 490, width=0.95\textwidth]{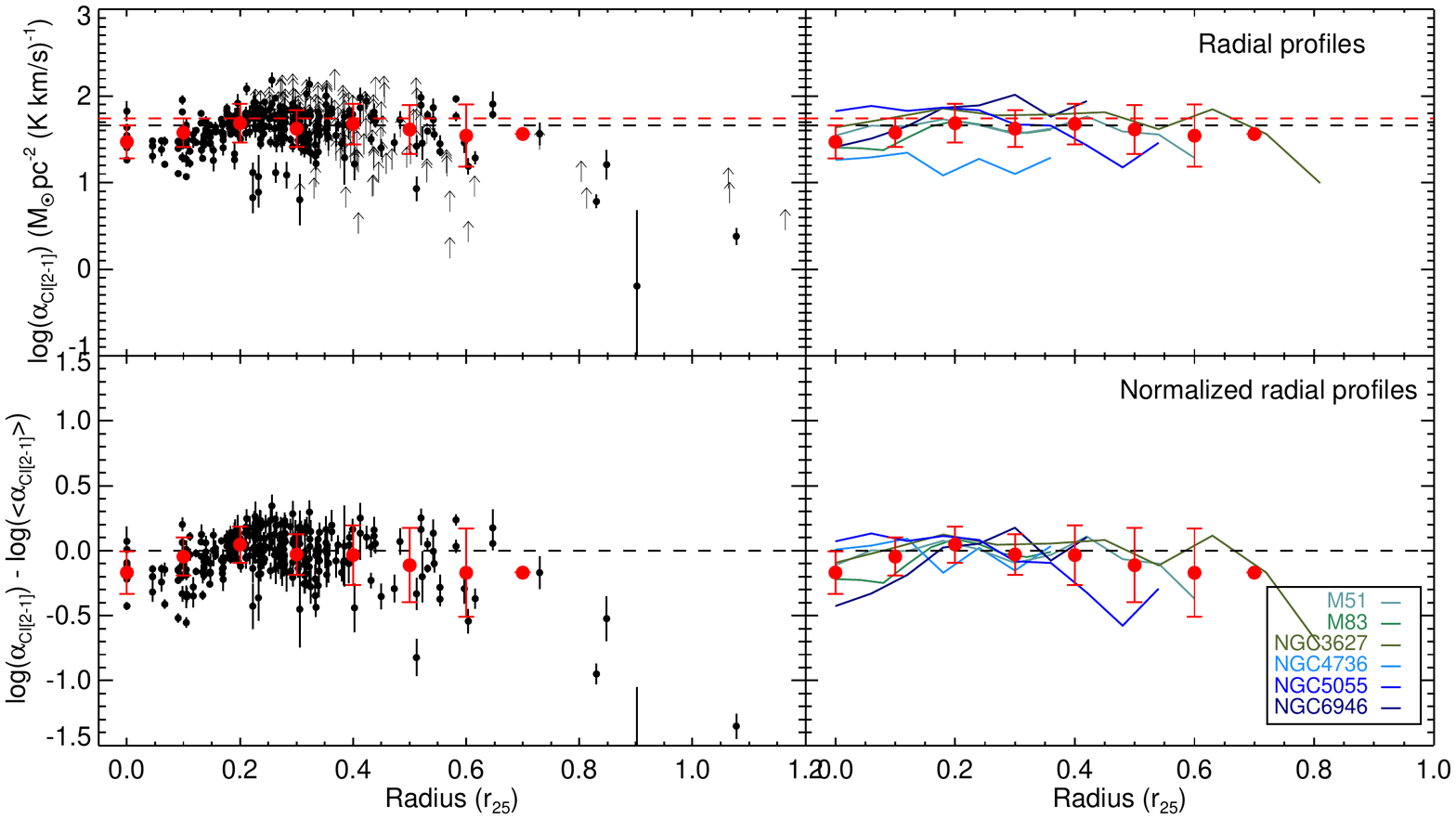}
 \end{center}
\caption{Same as Figure\,\ref{fig:aciwithradius_together} but for \alCtwo. The dotted black and red lines are the average values without (\alCtwo\ $= 46.1 \pm 21.8 M_{\odot }\,{\rm pc^{-2}\,(K\,km\,s^{-1})^{-1}}$) and with (\alCtwo\ $= 55.3 \pm 29.3 M_{\odot }\,{\rm pc^{-2}\,(K\,km\,s^{-1})^{-1}}$) non-detections for galaxies together, respectively.}
\label{fig:aci21withradius_together}
\end{figure*}

We present a summary of \alCO\ as a function of $r_{25}$ in Figure\,\ref{fig:acowithradius_together} with mean values shown as dotted lines, and 0.1$\,r_{25}$ bins as red symbols. In the right panels of Figure\,\ref{fig:acowithradius_together}, we also present the \alCO\ radial profiles of each galaxies with different colored lines. The top panels of Figure\,\ref{fig:acowithradius_together} show the estimated \alCO\ results and the bottom panels show the same values normalized by its galaxy-averaged \alCO. The radial profile of \alCO\ is generally flat as a function of $r_{25}$ with correlation coefficient $\rho =0.36$. But the correlation coefficient for the inner region becomes $\rho =0.42$ when limiting $r_{25}<0.3$, and the \alCO\ shows decrease in the inner region. As shown in bottom left panel of Figure\,\ref{fig:acowithradius_together}, the correlation coefficient between \alCO\ and galactocentric radii in the inner region ($r_{25}<0.3$) becomes a little more obvious with $\rho =0.49$ after normalizing each galaxy with its galaxy mean \alCO. In Table\,\ref{tab:alcicovalues}, we present the central and average values of \alCO\ for each galaxy. On average, for the \alCO\ in the same region as that of \CI\ observations, the central \alCO\ is $\sim 1.8$ times (ranging from $1.0-3.0$) lower than the galaxy average. And for the whole CO detection regions, the central values can be $\sim2.2$ times (ranging from $1.1-4.2$) lower than the galaxy means, on average.

Similarly, we present a summary of \alCone\ and \alCtwo\ as functions of $r_{25}$ in Figure\,\ref{fig:aciwithradius_together} and Figure\,\ref{fig:aci21withradius_together} with detections shown as filled black circles and non-detections as arrows. The red circles are 0.1$\,r_{25}$ bins of all detections, and the dotted black and red lines are the average values without and with non-detections for galaxies together, respectively.
 The central and average values of \alCone\ and  \alCtwo\ for each system are shown in Table\,\ref{tab:alcicovalues}. For the average $\alpha_{\rm [CI]}$ of each galaxy in Table\,\ref{tab:alcicovalues}, the first row is average value for the detections, and the second row shows the average value considering the non-detections with $enparCensored$ function. The detection points of \Cone\ for our sample are significantly smaller than \COone\ and \Ctwo, and the \alCone\ is primarily flat with galactocentric radius. As presented in Table\,\ref{tab:alcicovalues}, the central and average \alCone\ values are generally similar with each other for each galaxy. The \alCtwo\ is also mostly flat with galactocentric radius, while the central values of \alCtwo\ are slightly lower than the average values which can be seen more obvious in the bottom normalized panels of Figure\,\ref{fig:aci21withradius_together}. The central \alCtwo\ is on average $\sim 1.4$ times (ranging from $0.8-2.7$) lower than its galaxy average for our sample, and becomes $\sim 1.8$ times (ranging from $0.9 - 3.4$) lower when considering the limits.

For comparison, we also present the profiles of \alCone, \alCtwo, and \alCO\ for each galaxy when using the assumption with DGR$(ii)$ and DGR$(iii)$ in Figure\,\ref{fig:append_radii_DGRii},\,\ref{fig:append_radii_DGRiii}, and the summary profiles of galaxies together in Figure\,\ref{fig:acowithradius_together_DGR},\,\ref{fig:aciwithradius_together_DGR}, and \ref{fig:aci21withradius_together_DGR}. The profiles of \alCone, \alCtwo, and \alCO\ for each galaxy with different DGR assumptions look similar with each other, and different DGR assumptions only influence the specific values of the conversion factors. In Table\,\ref{tab:alcicovalues}, we present the central and average values of \alCone, \alCtwo, and \alCO\ for each system with different DGR assumptions. The values of \alCone, \alCtwo, and \alCO\ are similar for the assumptions of DGR$(i)$ and DGR$(ii)$, while smaller than that with the assumption of DGR$(iii)$. On average, the central values of \alCone, \alCtwo, \alCO\ in the \CI\ observation region, and \alCO\ in the whole galaxy detection region are $\sim$ 1.0, 1.4, 1.7, and 2.0 times lower than the galaxy averages of detections for both assumptions of DGR$(ii)$ and DGR$(iii)$, respectively. When the non-detections are considered for the galaxy averages, the central values of \alCone\ and \alCtwo\ are $\sim$ 1.1 and 1.6 times lower than the galaxy averages for the assumption of DGR$(ii)$, and become $\sim$ 1.1 and 1.7 times lower for the assumption of DGR$(iii)$, respectively.\\

\begin{table*}
  \caption{The central and average \alCO, \alCone, \alCtwo, and carbon abundance values for each sample}
  \label{tab:alcicovalues}
\begin{tabular}{lcccccccccc}
  \hline
\multirow{2}{*}{Name} &
\multicolumn{2}{c}{DGR$(i)$} &
\multicolumn{2}{c}{DGR$(ii)$} & 
\multicolumn{2}{c}{DGR$(iii)$} \\
%\cline{2-3}\cline{4-5}
 \cmidrule(r){2-3} \cmidrule(r){4-5} \cmidrule(r){6-7}
  & Central\_V$^a$ & Mean\_V$^b$ & 
  Central\_V & Mean\_V & Central\_V & Mean\_V & \multicolumn{2}{c}{} \\
\hline
\hline
\multicolumn{7}{c}{\alCO\ ($M_{\odot }\,{\rm pc^{-2}\,(K\,km\,s^{-1})^{-1}}$) in the same region as that of \CI\ observations}\\
\hline
      M~51 & $0.7 \pm 0.3$ & $0.9 \pm 0.4$ & $0.7 \pm 0.3$ & $0.9 \pm 0.4$ & $1.0 \pm 0.4$ & $1.2 \pm 0.6$ \\
      M~83 & $1.2 \pm 0.4$ & $2.0 \pm 0.8$ & $1.9 \pm 0.7$ & $2.8 \pm 1.0$ & $2.4 \pm 0.9$ & $3.5 \pm 1.3$ \\
      NGC~3627 & $0.8 \pm 0.3$ & $1.2 \pm 0.6$ & $1.0 \pm 0.4$ & $1.7 \pm 0.8$ & $1.2 \pm 0.4$ & $2.1 \pm 1.0$ \\
      NGC~4736 & $0.8 \pm 0.3$ & $2.0 \pm 2.1$ & $1.3 \pm 0.5$ & $3.4 \pm 3.1$ & $1.7 \pm 0.6$ & $4.6 \pm 4.1$\\
      NGC~5055 & $1.7 \pm 0.7$ & $ 1.7 \pm 0.3$ & $1.7 \pm 0.7$ & $ 1.7 \pm 0.3$ & $2.3 \pm 1.0$ & $2.5 \pm 0.4$ \\
      NGC~6946 & $0.6 \pm 0.2$ & $1.8 \pm 0.8$ & $1.3 \pm 0.5$ & $2.8 \pm 1.0$ & $1.6 \pm 0.6$ & $3.6 \pm 1.3$ \\
%      NGC~4826 & $9.20 \pm 0.04$ & ... \\
\hline
\hline
\multicolumn{7}{c}{\alCone\ ($M_{\odot }\,{\rm pc^{-2}\,(K\,km\,s^{-1})^{-1}}$)}\\
\hline
      M~51 & $12.8 \pm 1.1$ & $12.4 \pm 2.4$ & $12.8 \pm 1.1$ & $12.4 \pm 2.4$ & $17.3 \pm 1.4$ & $16.6 \pm 3.1$ \\
           & & $12.8 \pm 2.5$ & & $12.8 \pm 2.5$ & & $17.2 \pm 3.2$ \\
      M~83 & $11.8 \pm 0.7$ & $8.5 \pm 2.6$ & $19.0 \pm 1.1$ &  $13.1 \pm 4.1$ & $23.3 \pm 1.3$ & $16.1 \pm 5.0$ \\
           & & $11.8 \pm 3.9$ & & $17.0 \pm 4.9$ & & $21.0 \pm 6.1$ \\ 
      NGC~3627 & $14.9 \pm 0.7$ & $17.7 \pm 6.2 $ & $18.9 \pm 0.9$ & $23.4 \pm 8.0$ & $23.2 \pm 1.1$ & $29.5 \pm 10.1$\\
           & & $19.4 \pm 6.6$ & & $25.8 \pm 8.8$ & & $32.7 \pm 11.1$ \\
      NGC~4736 & $7.3 \pm 0.5$ & $7.3 \pm 2.7$ & $12.0 \pm 0.8$ & $12.7 \pm 4.6$ & $15.3 \pm 1.0$ & $16.8 \pm 6.1$ \\
           & & $8.0 \pm 1.8$ & & $14.0 \pm 3.0$ & & $18.7 \pm 4.1$ \\
      NGC~5055 & $16.7\pm 4.6$ & $ 12.8 \pm 2.8 $ & $16.7\pm 4.6$ & $ 12.8 \pm 2.8 $ & $22.9 \pm 6.0$ & $17.7 \pm 3.7$ \\
           & & $13.8 \pm 2.6$ & & $13.8 \pm 2.6$ & & $19.1 \pm 3.4$ \\
      NGC~6946 & $10.7 \pm 0.7$ & $13.0 \pm 3.7$ & $22.0 \pm 1.4$ & $23.3 \pm 6.3$ & $27.7 \pm 1.8$ & $29.4 \pm 7.9$ \\
           & & $14.8 \pm 4.4$ & & $25.2 \pm 6.8$ & & $32.1 \pm 8.7$ \\
\hline
\hline      
\multicolumn{7}{c}{\alCtwo\ ($M_{\odot }\,{\rm pc^{-2}\,(K\,km\,s^{-1})^{-1}}$)}\\
\hline
      M~51 & $35.0 \pm 2.5$ & $45.3 \pm 14.7$ & $35.0 \pm 2.5$ & $45.3 \pm 14.7$ & $47.2 \pm 3.3$ & $62.1 \pm 19.1$ \\
           & & $46.3 \pm 14.2$ & & $46.3 \pm 14.2$ & & $63.7 \pm 18.5$ \\
      M~83 & $25.4 \pm 1.6$ & $42.1 \pm 14.5$ & $40.9 \pm 2.6$ & $61.5 \pm 20.5$ & $50.3 \pm 3.1$ & $76.2 \pm 25.4$ \\
           & & $49.3 \pm 20.8$ & & $71.1 \pm 28.8$ & & $88.2 \pm 35.8$ \\
      NGC~3627 & $43.5 \pm 1.4$ & $53.9 \pm 25.5$ & $55.1 \pm 1.7$ & $70.1 \pm 35.3$ & $67.5 \pm 2.0$ & $88.9 \pm 44.6$  \\
           & & $54.6 \pm 24.7$ & & $71.3 \pm 34.0$ & & $90.3 \pm 43.0$ \\
      NGC~4736 & $18.3 \pm 0.9$ & $18.0 \pm 6.8$ & $30.0 \pm 1.1$ & $34.4 \pm 9.8$ & $38.2 \pm 1.2$ & $47.9 \pm 13.0$ \\
           & & $32.2 \pm 19.6$ & & $51.1 \pm 25.1$ & & $68.1 \pm 31.0$ \\
      NGC~5055 & $67.0\pm 18.1$ & $ 56.8 \pm 21.1$ & $67.0\pm 18.1$ & $ 56.8 \pm 21.1$ & $91.9 \pm 23.9$ & $77.6 \pm 31.5$ \\
           & & $62.7 \pm 20.2$ & & $62.7 \pm 20.2$ & & $86.9 \pm 29.8$ \\ 
      NGC~6946 & $25.8 \pm 1.9$ & $69.0 \pm 32.2$ & $53.1 \pm 3.7$ & $119.3 \pm 52.1$ & $67.0 \pm 4.7$ & $151.3 \pm 66.2$ \\
           & & $88.6 \pm 37.6$ & & $144.9 \pm 59.3$ & & $184.2 \pm 75.3$ \\
\hline
\hline      
\multicolumn{7}{c}{Carbon abundance ($\times 10^{-5}$)}\\
\hline
      M~51 & $1.7 \pm 0.2$ & $1.9 \pm 0.4$ & $1.7 \pm 0.2$ & $1.9 \pm 0.4$ & $1.2 \pm 0.1$ & $1.4 \pm 0.3$ \\
      M~83 & $1.8 \pm 0.1$ & $3.0 \pm 1.3$ & $1.1 \pm 0.1$ & $2.0 \pm 0.8$ & $0.9 \pm 0.1$ & $1.6 \pm 0.7$ \\
      NGC~3627 & $1.5 \pm 0.9$ & $1.4 \pm 0.6$ & $1.1 \pm 0.1$ & $1.1 \pm 0.4$ & $0.9 \pm 0.1$ & $0.9 \pm 0.3$ \\
      NGC~4736 & $2.9 \pm 0.3$ & $2.7 \pm 0.5$ & $1.8 \pm 0.1$ & $1.6 \pm 0.3$ & $1.4 \pm 0.1$ & $1.2 \pm 0.2$ \\
      NGC~5055 & $1.4 \pm 0.4$ & $2.0 \pm 0.5 $ & $1.4 \pm 0.4$ & $2.0 \pm 0.5 $ & $1.0 \pm 0.3$ & $1.5 \pm 0.3$ \\
      NGC~6946 & $2.0 \pm 0.2$ & $1.9 \pm 0.5$ & $1.0 \pm 0.1$ & $1.1 \pm 0.3$ & $0.8 \pm 0.1$ & $0.8 \pm 0.2$ \\
\hline 
\hline      
\multicolumn{7}{c}{\alCO\ ($M_{\odot }\,{\rm pc^{-2}\,(K\,km\,s^{-1})^{-1}}$) for the whole CO detection region}\\
\hline
      M~51 & $0.7 \pm 0.3$ & $1.4 \pm 1.3$ & $0.7 \pm 0.3$ & $1.4 \pm 1.3$ & $1.0 \pm 0.4$ & $2.0 \pm 1.7$ \\
      M~83 & $1.2 \pm 0.4$ & $2.4 \pm 1.1$ & $1.9 \pm 0.7$ & $3.3 \pm 1.3$ & $2.4 \pm 0.9$ & $4.1 \pm 1.6$ \\
      NGC~3627 & $0.8 \pm 0.3$ & $1.2 \pm 0.7$ & $1.0 \pm 0.4$  & $1.7 \pm 1.0$ & $1.2 \pm 0.4$ & $2.2 \pm 1.2$ \\
      NGC~4736 & $0.8 \pm 0.3$ & $2.0 \pm 2.1$ & $1.3 \pm 0.5$ & $3.4 \pm 3.1$ & $1.7 \pm 0.6$ & $4.6 \pm 4.1$ \\
      NGC~5055 & $1.7 \pm 0.7$ & $1.8 \pm 0.3$ & $1.7 \pm 0.7$ & $1.8 \pm 0.3$ & $2.3 \pm 1.0$ & $2.5 \pm 0.4$ \\
      NGC~6946 & $0.6 \pm 0.2$ & $2.5 \pm 1.2$ & $1.3 \pm 0.5$ & $3.4 \pm 1.2$ & $1.6 \pm 0.6$ & $4.3 \pm 1.6$ \\
\hline       
%\multicolumn{7}{l}{$^a$ Central value of each galaxy.}\\
%\multicolumn{7}{l}{$^b$ Average value of each galaxy. }\\
\end{tabular}\\
\footnotesize{$^{\rm a}$ Central value of each galaxy.}\\
\footnotesize{$^{\rm b}$ Average value of each galaxy. For $\alpha_{\rm [CI]}$ of each galaxy, the first row is average value for the detections, and the second row shows the average value which takes into consideration the non-detections using $enparCensored$ function.}\\
\end{table*}

\subsection{\CI\ abundance}

Under optically thin and local thermodynamical equilibrium (LTE) assumptions, the atomic carbon mass can be derived using:
\begin{eqnarray}
 M_\mathrm{[CI]} & = &C m_\mathrm{[CI]} \frac {8\pi k \nu_0^2}{hc^2A_{\rm 10}} Q(T_{\rm ex}) \frac {1}{3} e^{T_1/T_{\rm ex}}{L'_\mathrm{[CI](1-0)}}\nonumber \\
      &=& 5.706\times 10^{-4}Q(T_{\rm ex})\frac {1}{3} e^{23.6/T_{\rm ex}}{L'_\mathrm{[CI](1-0)}},      
\label{equ_massCI10}    
\end{eqnarray}
% or
%\begin{eqnarray}
% M_\mathrm{[CI]} &=&C m_\mathrm{[CI]} \frac {8\pi k \nu_0^2}{hc^2A_{\rm 21}} Q(T_{\rm ex}) \frac {1}{5} e^{T_2/T_{\rm ex}}{L'_\mathrm{[CI](2-1)}} \nonumber \\
%     &=&4.566\times 10^{-4}Q(T_{\rm ex})\frac {1}{5} e^{62.5/T_{\rm ex}}{L'_\mathrm{[CI](2-1)}},    
%\label{equ_massCI21}     
%\end{eqnarray}
with \Cone\ luminosities \citep{Weiss et al. 2003, Weiss et al. 2005}. Among the equation, $C$ is the conversion between $\mathrm{pc^2}$ to $\mathrm{cm^2}$, $m_\mathrm{[CI]}$ represents the atomic carbon mass, and $A_\mathrm{10}=7.93 \times 10^{-8}\, \mathrm{s^{-1}}$ is the Einstein coefficient. $T_\mathrm{ex}$ is the \CI\ excitation temperature which can be estimated using $T_\mathrm{ex} = 38.8\, \mathrm{K/ln[2.11}/R_\mathrm{[CI]}]$ under optically thin condition \citep{Stutzki et al. 1997} with $R_\mathrm{[CI]}=L'_\mathrm{[CI](2-1)}/L'_\mathrm{[CI](1-0)}$. $Q_{\rm ex}= 1 + 3e^{-T_1/T_{\rm ex}} + 5e^{-T_2/T_{\rm ex}}$ is the \CI\ partition function which depends on excitation temperature $T_{\rm ex}$ with $T_1=23.6\ $K and $T_2=62.5\ $K (the energies above the ground state). The details of $T_\mathrm{ex}$ for each galaxy can be found in \citet{Jiao et al. 2019}. %Thus carbon mass is a function of $T_{\rm ex}$ and \CI\ luminosities under optically thin and LTE conditions.

\begin{figure*}
 \begin{center}
  \includegraphics[bb = 67 115 730 490, width=0.95\textwidth]{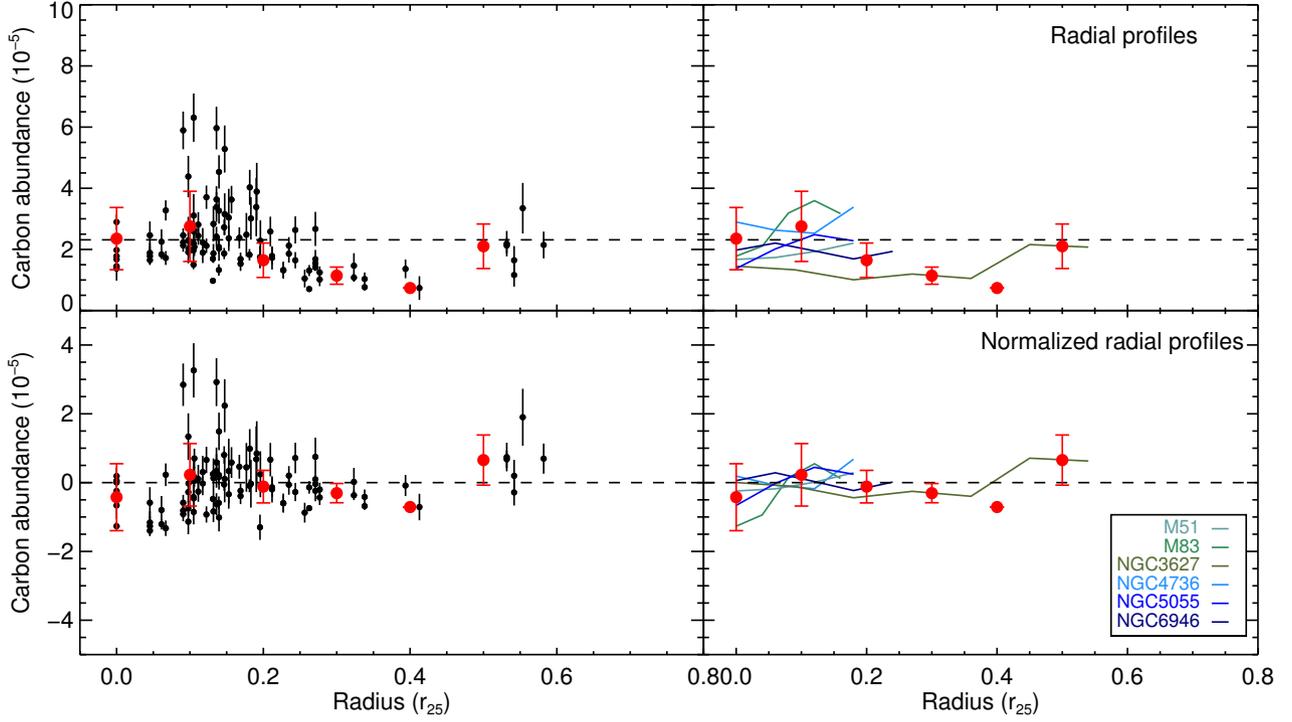} 
 \end{center}
\caption{The left panels show the carbon abundance for detections of each galaxy together as a functions of galactocentric radii. The right panels show the radial profile for each galaxy. The top and bottom panels show the original and normalized carbon abundances. The filled black circles represent each points for the whole sample, and the mean and standard deviation of all points in 0.1$\,r_{25}$ bins are shown with red symbols. }\label{fig:ciabundancewithradius_together}
\end{figure*}

The H$_2$ mass can be derived with Equation\,\ref{equ_estimate_method}, and then we can obtain the carbon abundance using mass ratio between \CI\ and H$_2$: $X[{\rm CI}]/X[{\rm H_2}]=M({\rm [CI]})/(6M({\rm H_2}))$. Using the assumption of DGR$(i)$, we present a summary of carbon abundance as a function of $r_{25}$ with mean values shown as dotted lines in Figure\,\ref{fig:ciabundancewithradius_together}. We also list the central and  average values of carbon abundance for each galaxy in Table\,\ref{tab:alcicovalues}. The scatter in Figure\,\ref{fig:ciabundancewithradius_together} is dramatical, and we cannot find obvious correlation between carbon abundance with $r_{25}$. The central and average carbon abundances for each system are comparable. The average carbon abundance of the sample is $2.3 \pm 1.1 \times 10^{-5}$, which is comparable with the commonly adopted abundance of $X[{\rm CI}]/X[{\rm H_2}] \sim 3.0 \times 10^{-5}$ \citep{Weiss et al. 2003, Papadopoulos et al. 2004amodel}.

We also present the carbon abundances estimated with assumptions of DGR$(ii)$ and DGR$(iii)$ in Figure\,\ref{fig:ciabundancewithradius_together_DGR} and Table\,\ref{tab:alcicovalues}. The profiles of carbon abundances with different DGR assumptions look similar with each other. The average carbon abundance with assumption of DGR$(ii)$ ($X[{\rm CI}]/X[{\rm H_2}] \sim 1.6 \pm 0.7 \times 10^{-5}$) is comparable with the value when using the assumption of DGR$(iii)$ ($X[{\rm CI}]/X[{\rm H_2}] \sim 1.2 \pm 0.6 \times 10^{-5}$).

\subsection{Correlations with Environmental Parameters}

\begin{table*}
  \caption{The correlation properties of \alCO, \alCone, \alCtwo, and carbon abundance.}
  \label{tab:coefficient}
  \begin{tabular}{lccccccccccc}
  \hline
\multirow{2}{*}{Varable vs.} &
\multicolumn{2}{c}{DGR$(i)$} &
\multicolumn{2}{c}{DGR$(ii)$} & 
\multicolumn{2}{c}{DGR$(iii)$} \\
%\cline{2-3}\cline{4-5}
 \cmidrule(r){2-3} \cmidrule(r){4-5} \cmidrule(r){6-7}
  & $\rho$ & p-value & 
  $\rho$ & p-value & $\rho$ & p-value & \multicolumn{2}{c}{} \\
\hline
\hline
\multicolumn{7}{c}{\alCO\ in the same region as that of \CI\ observations }\\
\hline 
      $\overline{U}$ & $-0.18$ & $5.11\times10^{-4}$ & $-0.04$ & $0.43$ & $-0.09$ & $0.08$ \\
      ${\rm log}(L_{\rm IR})$ & $-0.57$ & $3.31\times 10^{-32}$ & $-0.64$ & $4.96\times 10^{-42}$ & $-0.62$ & $1.20\times 10^{-39}$ \\
      $12 + {\rm log(O/H)}$ & $-0.63$ & $7.27\times 10^{-41}$ & $-0.75$ & $\sim 0$ & $-0.74$ & $\sim 0$ \\
\hline
\hline
\multicolumn{7}{c}{\alCone}\\
\hline
      $\overline{U}$ & $-0.42$ & $9.53\times10^{-5}$ & $-0.19$ & $0.09$ & $-0.23$ & $0.04$\\
      ${\rm log}(L_{\rm IR})$ & $0.58$ & $1.92 \times 10^{-8}$ & $0.41$ & $1.47 \times 10^{-4}$ & $0.47$ & $7.51 \times 10^{-6}$ \\
      $12 + {\rm log(O/H)}$ & $0.59$ & $5.18 \times 10^{-9}$ & $0.31$ & $4.73 \times 10^{-3}$ & $0.36$ & $1.07 \times 10^{-3}$ \\
\hline
\hline      
\multicolumn{7}{c}{\alCtwo}\\
\hline
      $\overline{U}$ & $-0.49$ & $3.40 \times 10^{-17}$ & $-0.28$ & $3.10 \times 10^{-6}$ & $-0.34$ & $9.45 \times 10^{-9}$  \\
      ${\rm log}(L_{\rm IR})$ & $0.16$ & $0.01$ & $-0.02$ & $0.77$ & $0.05$ & $0.41$ \\
      $12 + {\rm log(O/H)}$ & $0.22$ & $3.85\times10^{-4}$ & $-0.01$ & $0.84$ & $0.06$ & $0.33$ \\
\hline
\hline      
\multicolumn{7}{c}{Carbon abundance ($\times 10^{-5}$)}\\
\hline
      $\overline{U}$ & $0.32$ & $4.57\times10^{-3}$ & $0.15$ & $0.20$ & $0.18$ & $0.13$ \\
      ${\rm log}(L_{\rm IR})$ & $-0.59$ & $1.27 \times 10^{-8}$ & $-0.43$ & $9.58 \times 10^{-5}$ & $-0.47$ & $1.30 \times 10^{-5}$\\
      $12 + {\rm log(O/H)}$ & $-0.58$ & $3.92 \times 10^{-8}$ & $-0.30$ & $7.14 \times 10^{-3}$ & $-0.33$ & $3.27 \times 10^{-3}$ \\
\hline       
\end{tabular}
\end{table*}

\begin{figure*}
 \begin{center}
  \includegraphics[bb= 50 230 740 440, width=0.93\textwidth]{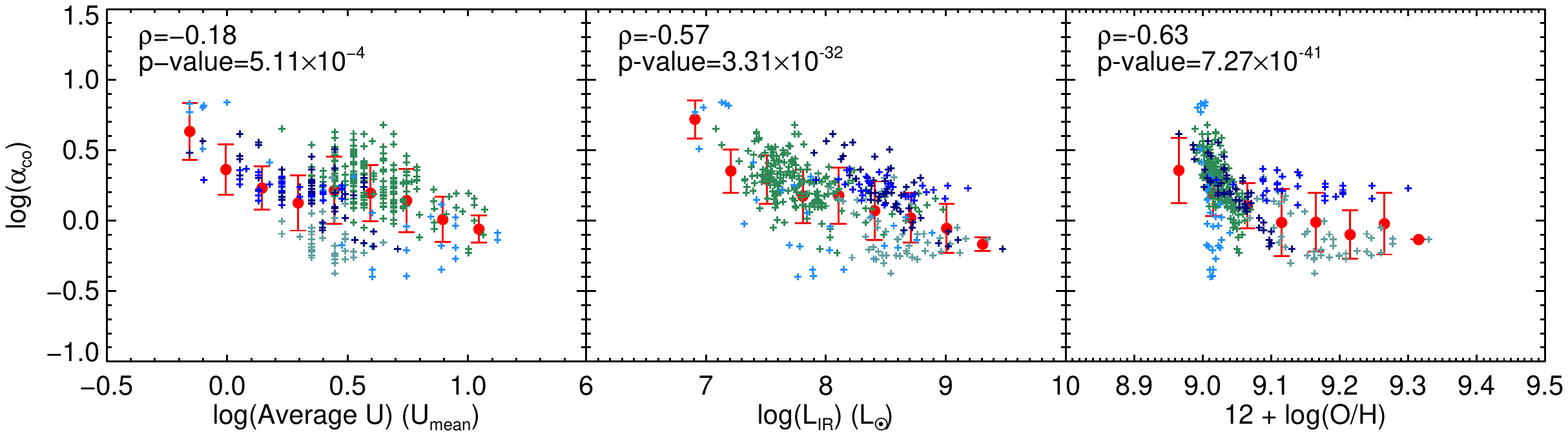}\\
  \includegraphics[bb= 50 230 740 440, width=0.93\textwidth]{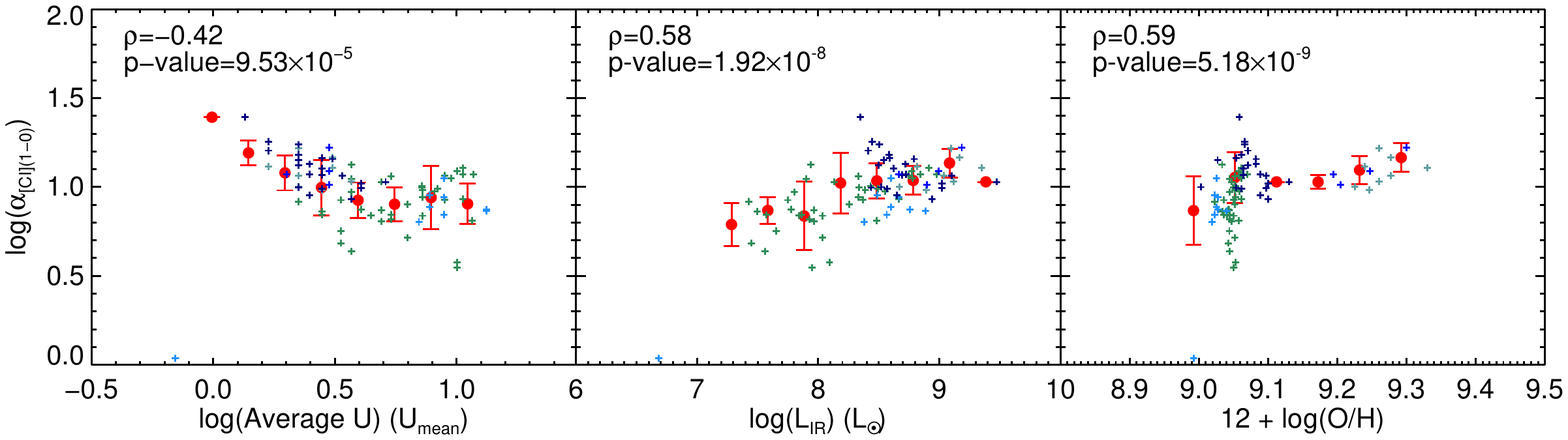} \\
  \includegraphics[bb= 50 230 740 440, width=0.93\textwidth]{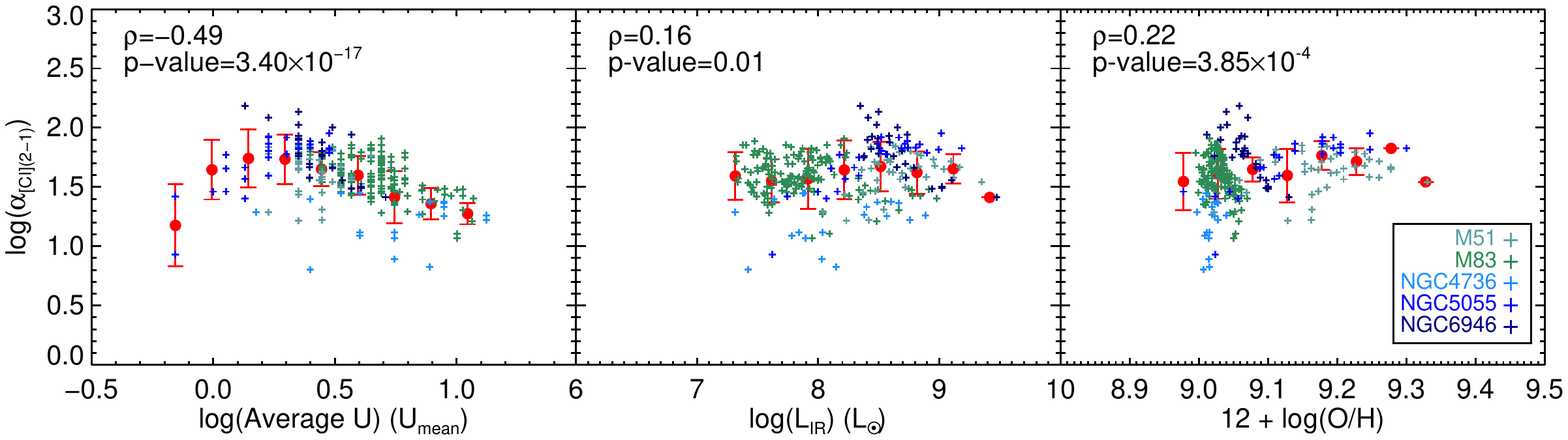}
 \end{center}
\caption{\alCO\ (top panels), \alCone\ (middle panels), \alCtwo\ (bottom panels) as functions of environmental parameters, i.e., $\overline{U}$ (first column), $log(L_{\rm IR})$ (second column), and metallicity (third column). The colored plus signs represent each galaxy detections, and the red symbols show the 0.1$\,\rm {r_{25}}$ bin values of all detections. The labeled $\rho$ and p-value represent the correlation coefficient and the possibility of no correlation.} %The black line in the right-middle panel presents the best-fit linear relation log$\,$\alCone\ $= -1.13\times {\rm log(Z/Z_{\sun})} + 1.33$ from \citet{Heintz Watson 2020}.} 
\label{fig:acociwithparas}
\end{figure*}

\begin{figure*}
 \begin{center}
  \includegraphics[bb= 50 230 740 440, width=0.93\textwidth]{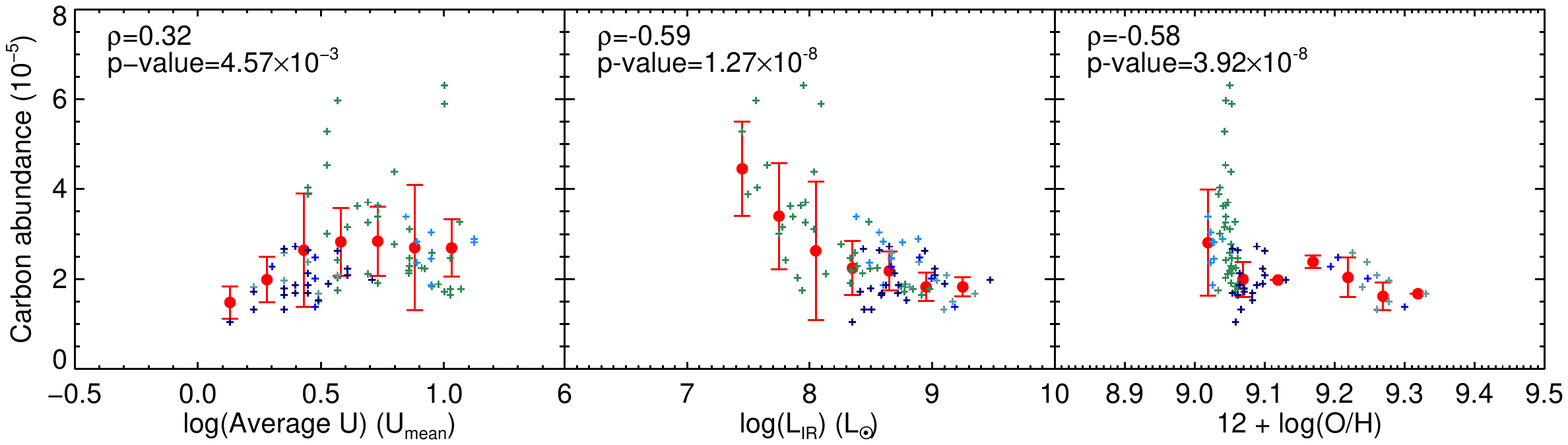} \\
  \includegraphics[bb= 50 230 740 440, width=0.93\textwidth]{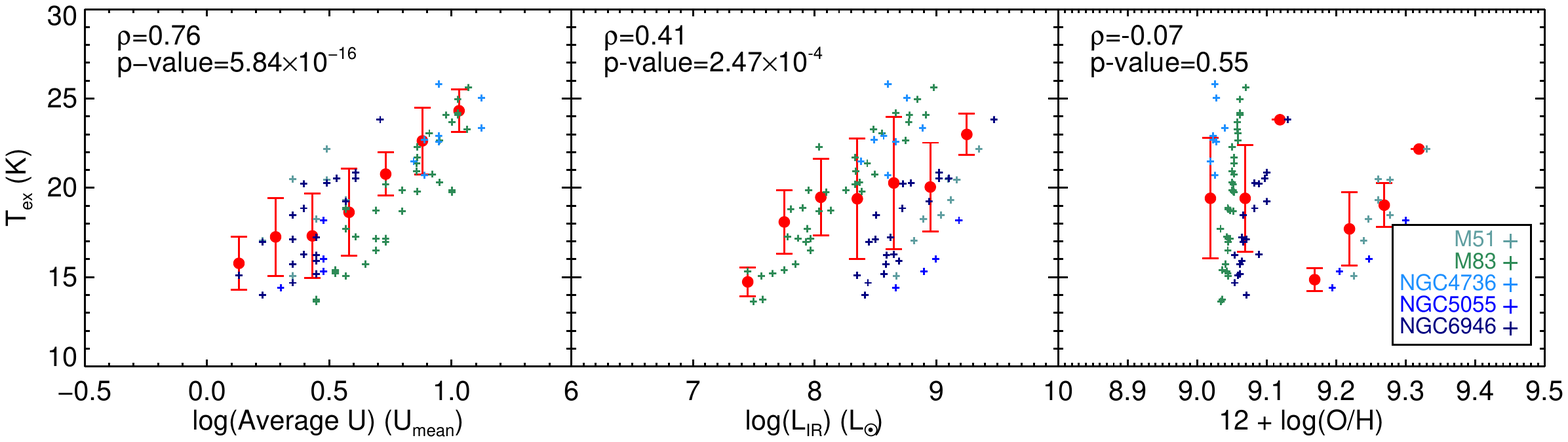} 
 \end{center}
\caption{Carbon abundance and $T_\mathrm{ex}$ as functions of $\overline{U}$ (first column), $log(L_{\rm IR})$ (second column), and metallicity (third column). The colored plus signs represent each galaxy detections, and the red symbols show the 0.1$\,\rm {r_{25}}$ bin values of all detections. The labeled $\rho$ and p-value represent the correlation coefficient and the possibility of no correlation.}\label{fig:abundancewithparas}
\end{figure*}

\begin{figure}
 \begin{center}
  \includegraphics[bb= 135 80 580 490, width=0.47\textwidth]{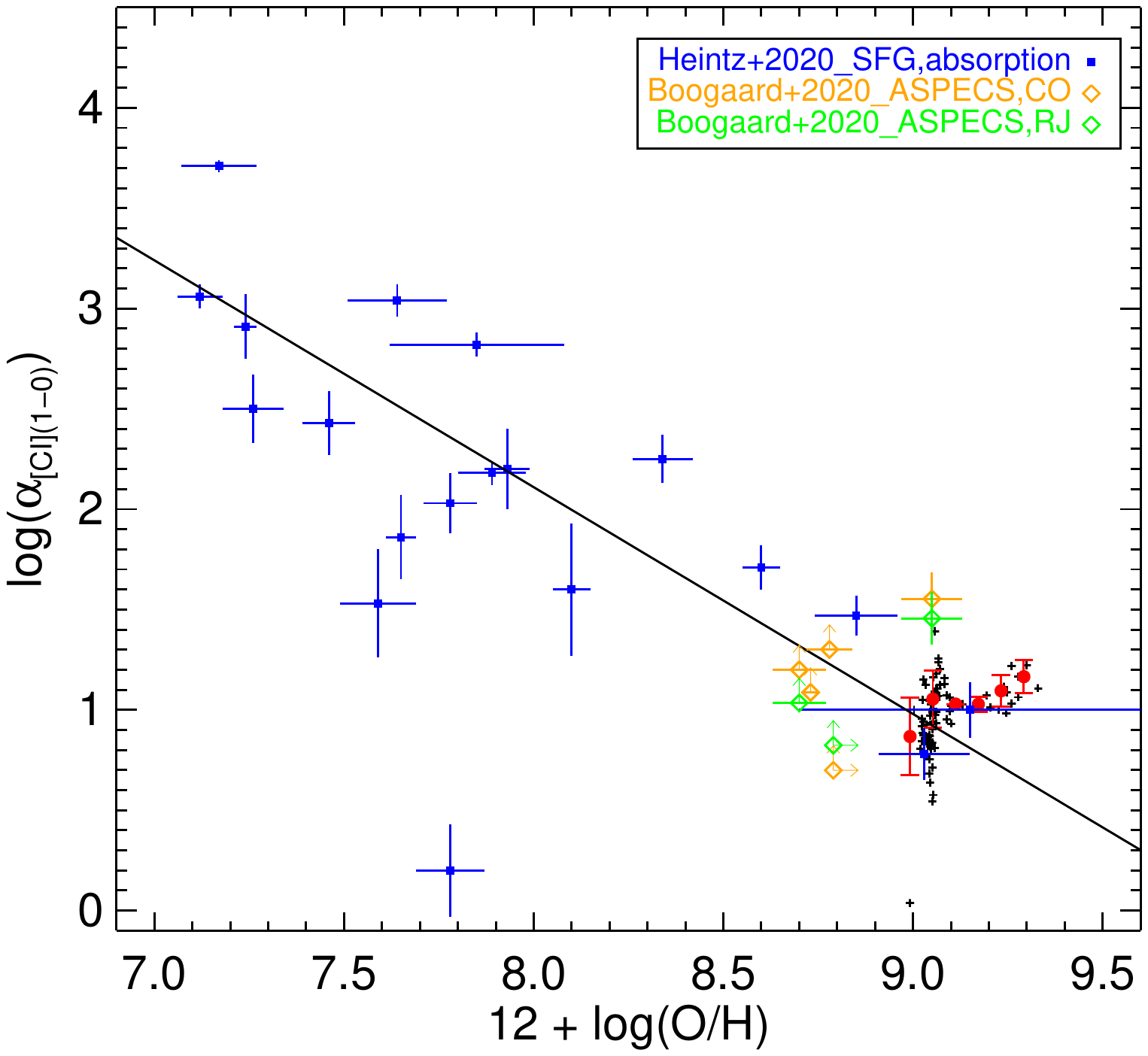} 
 \end{center}
\caption{\alCone\ as a function of metallicity. The black and red symbols represent each detections and 0.1$\,\rm {r_{25}}$ bin values  for our sample. The blue boxes represent samples of high-redshift ($z = 1.9 - 3.4$) gamma-ray burst and quasar molecular gas absorbers from \citet{Heintz Watson 2020} with $12 + \rm{log(O/H)} \sim 7.12 - 9.15$, and the black line shows their best-fit linear relation of log$\,$\alCone\ $= -1.13\times {\rm log(Z/Z_{\sun})} + 1.33$. The orange and green diamonds are ASPECS galaxies from \citet{Boogaard et al. 2020} with molecular gas masses estimated based on CO luminosity and 1.2 mm dust-continuum emission on the Rayleigh-Jeans tail, respectively.}\label{fig:acociwithheniz2020}
\end{figure}

In the following analysis, we exclude galaxy NGC~3627 which has no available metallicity gradient from  \citet{Moustakas et al. 2010}. In Figure\,\ref{fig:acociwithparas}, we plot the \alCO, \alCone, and \alCtwo\ values as functions of different  physical properties of galaxies, i.e., the average interstellar radiation field $\overline{U}$, the infrared luminosity ${\rm log}(L_{\rm IR})$, the metallicity $12 + {\rm log(O/H)}$. The labeled $\rho$ and p-value in each panel represent the correlation coefficient and the possibility of no correlation.%, and the red symbols show the 0.1$\,r_{25}$ bins. 

The top panels of Figure\,\ref{fig:acociwithparas} show that \alCO\ has no correlation with $\overline{U}$, while correlates moderately and decreases with both ${\rm log}(L_{\rm IR})$ and metallicity of  $12 + {\rm log(O/H)}$. %\alCone\ has no correlation with $\overline{U}$, and correlates weakly with  ${\rm log}(L_{\rm IR})$ and metallicity $12 + {\rm log(O/H)}$ with coefficient of $\rho=0.38$, and $\rho=0.33$, respectively. 
\alCone\ correlates weakly with $\overline{U}$, ${\rm log}(L_{\rm IR})$, and metallicity. \alCtwo\ only has weak correlation with $\overline{U}$, and has no obvious correlation with ${\rm log}(L_{\rm IR})$ and metallicity $12 + {\rm log(O/H)}$. Compared to flat correlation between \alCone\ with metallicity in almost all other galaxies as shown in the right column of Figure\,\ref{fig:acociwithparas}, the \alCone\ of M~83 changes dramatically with metallicity with $\rho = 0.40$. With the largest detection number in our sample, M~83 may significantly impact the final results. So we also estimate correlation coefficients without M~83. And the correlation between \alCone\ with metallicity becomes $\rho = 0.38$ when M~83 is not taken into consideration. The strong correlation of M~83 compared to other galaxies might be due to its intense starburst than other milder starburst of NGC~6946 and AGNs in our sample, which enhances the carbon excitation and leads to a higher neutral carbon to CO column density ratio \citep{Israel & Baas 2001, Jiao et al. 2019}. Besides, as also shown in \citet{Jiao et al. 2019}, the linear correlations between \LCOone\ with both $L'_{\rm [CI]}$ of M~83 are steeper than other galaxies in their sample. However, we also need to note that the metallicity calibration method of M~83 is different with other galaxies in our sample, and M~83 mainly locates around $12 + {\rm log(O/H)} \sim 9.03$ which is smaller than most of other sample galaxies.

\citet{Sandstrom et al. 2013} found that the \alCO\ for their sample has no obvious correlation with average interstellar radiation field, which is constant with our result. While they also found no obvious correlations between \alCO\ with metallicity and star-formation rate surface density estimated from H$\alpha$ and 24\,\mum\ maps. The $L_{\rm IR}$ has been widely used as an indicator of star-formation rate (SFR) in galaxies \citep{Kennicutt Evans 2012}, and thus the good correlation between \alCO\ and ${\rm log}(L_{\rm IR})$ in our sample is inconstant with \citet{Sandstrom et al. 2013}. However, the \alCO\ in central starburst region in the galaxies of \citet{Sandstrom et al. 2013} shows two times below the galaxy mean on average. Lower \alCO\ has been found in starburst galaxies (e.g., \citealt{Mao et al. 2000, Hinz & Rieke 2006,  Zhu et al. 2009, Meier et al 2010, Cormier et al. 2018}), interacting systems \citep{Gao et al. 2001, Gao et al. 2003, Zhu et al. 2003, Zhu et al. 2007}, and LIRGs with extreme star formation activities \citep{Downes Solomon 1998, Kamenetzky et al. 2014, Sliwa et al. 2017}. \citet{Narayanan et al. 2011} derived that the \alCO\ drops by a typical factor of $\sim 2-10$ throughout the actively star-forming area in starbursts with hydrodynamic simulations of disk and merging galaxies, and they attributed the lower \alCO\ to higher gas temperatures and very large velocity dispersions. Thus the \alCO\ drops in massive mergers during the starburst phase, with low \alCO\ corresponding to high peak SFR, and settles to normal values when the star formation activity and the conditions that caused it subside (see  \citealt{Narayanan et al. 2011, Bolatto et al. 2013}).
But we also need to take care that most of our six galaxies are AGNs, and emission from AGN can also heat dust and make significant contribution to IR luminosity for the central regions which may overestimate the true SFR \citep{Hayward et al. 2014, Dai et al. 2018, Hickox & Alexander 2018}. Many theoretical and observational studies have shown that metallicity is an important driver for \alCO\ variations (e.g., \citealt{Leroy et al. 2011, Feldmann et al. 2012, Narayanan et al. 2012, Bolatto et al. 2013}), which agree well with our result. 

We also present carbon abundance and $T_\mathrm{ex}$ as functions of $\overline{U}$, ${\rm log}(L_{\rm IR})$, and metallicity $12 + {\rm log(O/H)}$ in Figure\,\ref{fig:abundancewithparas}. The carbon abundance shows weak correlation with $\overline{U}$, ${\rm log}(L_{\rm IR})$ and metallicity. And $T_\mathrm{ex}$ shows moderate and weak correlation with $\overline{U}$ and ${\rm log}(L_{\rm IR})$, and has no correlation with metallicity. In Table\,\ref{tab:coefficient}, we list the correlation coefficients and the possibilities of no correlation between CO,  \CI\ conversion factors, and carbon abundance with environmental parameters using different DGR assumptions. And the correlation coefficients for different DGR assumptions agree with each other well.

\subsection{Comparison to the literature and discussion}

\citet{Offner et al. 2014} used 3{\scriptsize D-PDR} to post-process hydrodynamic simulation of turbulent star-forming clouds, and derived an average `$X$-factor\footnote{Without considering the helium and heavier elements, the \CI-to-H$_2$ and CO-to-H$_2$ conversion factors can be converted from unit of ${\rm cm^{-2}\,(K\,km\,s^{-1})^{-1}}$ to unit of $M_{\odot }\,{\rm pc^{-2}\,(K\,km\,s^{-1})^{-1}}$ by multiplying a factor of $1.6 \times 10^{-20}$, i.e., \alCO\,$M_{\odot }\,{\rm pc^{-2}\,(K\,km\,s^{-1})^{-1}} = 1.6 \times 10^{-20} \, X_{\rm CO}\,{\rm cm^{-2}\,(K\,km\,s^{-1})^{-1}}$, and $\alpha_{\rm [CI]}\,M_{\odot }\,{\rm pc^{-2}\,(K\,km\,s^{-1})^{-1}} = 1.6 \times 10^{-20}\, X_{\rm [CI]} \,{\rm cm^{-2}\,(K\,km\,s^{-1})^{-1}}$. And the factor becomes $1.6 \times 10^{-20} \times 1.36 = 2.2 \times 10^{-20}$ when including the helium and heavier elements. We use the \CI\ and CO conversion factors in mass unit without helium and heavier elements corrections throughout the paper.}' of $X_{\rm CO} = 3.0 \times 10^{20}\,{\rm cm^{-2}\,(K\,km\,s^{-1})^{-1}}$ and $X_{\rm [CI](1-0)} = 1.1 \times 10^{21}\,{\rm cm^{-2}\,(K\,km\,s^{-1})^{-1}}$. These values correspond to \alCO\ $=4.8$ $M_{\odot }\,{\rm pc^{-2}\,(K\,km\,s^{-1})^{-1}}$ and \alCone\ $= 17.6$ $M_{\odot }\,{\rm pc^{-2}\,(K\,km\,s^{-1})^{-1}}$.  By utilizing a modified astrochemistry code that includes different cosmic rays which stand for extreme, star-forming, and quiescent regions, \citet{Gaches et al. 2019} derived a $X_{\rm [CI](1-0)}$ ranging from $2 \times 10^{20} < X_{\rm [CI](1-0)} < 4 \times 10^{21}\,{\rm cm^{-2}\,(K\,km\,s^{-1})^{-1}}$ (corresponding to $3.2 < \alpha_{\rm [CI](1-0)} < 64.1\,M_{\odot }\,{\rm pc^{-2}\,(K\,km\,s^{-1})^{-1}}$). Our obtained \alCone\ values are comparable with both works, and \alCO\ values are smaller than the result of \citet{Offner et al. 2014}.

\citet{Israel 2020} collected the central \CI\ line data from \citet{Lu et al. 2017}, \citet{Israel et al. 2015}, and \citet{Kamenetzky et al. 2016} which were mostly obtained from $Herschel$, and then reduced these \CI\ fluxes to their ``standard'' beam size of 22$"$ with 35$"$ to 22$"$ beam conversion factors. Using the beam corrected \CI\ fluxes and CO data observed with ground-based measurements, \citet{Israel 2020} then obtained an average $X_{\rm CO} = 1.9 \times 10^{19}\,{\rm cm^{-2}\,(K\,km\,s^{-1})^{-1}}$ and $X_{\rm [CI](1-0)} = 9.1 \times 10^{19}\,{\rm cm^{-2}\,(K\,km\,s^{-1})^{-1}}$ which is corresponding to \alCO\ $=0.3$ $M_{\odot }\,{\rm pc^{-2}\,(K\,km\,s^{-1})^{-1}}$ and \alCone\ $= 1.4$ $M_{\odot }\,{\rm pc^{-2}\,(K\,km\,s^{-1})^{-1}}$ for a sample of nearby galaxy centers with molecular hydrogen column densities estimated based on the statistical equilibrium radiative transfer code RADEX \citep{Van der Tak et al. 2007} and carbon abundance. The average conversion factors of \alCone\ $= 3.6$ $M_{\odot }\,{\rm pc^{-2}\,(K\,km\,s^{-1})^{-1}}$ and \alCtwo\ $= 12.5$ $M_{\odot }\,{\rm pc^{-2}\,(K\,km\,s^{-1})^{-1}}$ for (U)LIRGs \citep{Jiao et al. 2017}, and \alCone\ $= 7.3$ $M_{\odot }\,{\rm pc^{-2}\,(K\,km\,s^{-1})^{-1}}$ and \alCtwo\ $= 34$ $M_{\odot }\,{\rm pc^{-2}\,(K\,km\,s^{-1})^{-1}}$ for 18 nearby galaxies \citep{Crocker et al. 2019} are smaller than our results. \citet{Izumi et al. 2020} and \citet{Miyamoto et al. 2021} found lower \alCone\ in the centre of NGC~7469 (\alCone\ $= 4.4$ $M_{\odot }\,{\rm pc^{-2}\,(K\,km\,s^{-1})^{-1}}$) and northern part of M~83 (\alCone\ $= 3.8$ $M_{\odot }\,{\rm pc^{-2}\,(K\,km\,s^{-1})^{-1}}$), respectively. It must be note that the $\alpha_{\rm [CI]}$ in \citet{Jiao et al. 2017} and \citet{Crocker et al. 2019} are estimated by adopting an assumed \alCO.  \alCone\ in \citet{Izumi et al. 2020} is based on dynamical modelings, and in \citet{Miyamoto et al. 2021} is estimated with assumptions of gas-to-dust ratio.
%\citet{Jiao et al. 2017} and \citet{Crocker et al. 2019} obtained average values of \alCone\ $= 3.6$ $M_{\odot }\,{\rm pc^{-2}\,(K\,km\,s^{-1})^{-1}}$ and \alCtwo\ $= 12.5$ $M_{\odot }\,{\rm pc^{-2}\,(K\,km\,s^{-1})^{-1}}$ for (U)LIRGs, and \alCone\ $= 7.3$ $M_{\odot }\,{\rm pc^{-2}\,(K\,km\,s^{-1})^{-1}}$ and \alCtwo\ $= 34$ $M_{\odot }\,{\rm pc^{-2}\,(K\,km\,s^{-1})^{-1}}$ for 18 nearby galaxies by adopting an assumed \alCO, respectively.  \citet{Crocker et al. 2019} derived mean values of \alCone\ $= 7.3$ $M_{\odot }\,{\rm pc^{-2}\,(K\,km\,s^{-1})^{-1}}$ and \alCtwo\ $= 34$ $M_{\odot }\,{\rm pc^{-2}\,(K\,km\,s^{-1})^{-1}}$ for 18 nearby galaxies observed with the $Herschel$ using CO\,(2$-$1) and \alCO\ results from \citet{Sandstrom et al. 2013}. Adopting \alCO\ $= 0.8$ $M_{\odot }\,{\rm pc^{-2}\,(K\,km\,s^{-1})^{-1}}$ for a sample of nearby (U)LIRGs, \citet{Jiao et al. 2017} obtained average results of \alCone\ $= 3.6$ $M_{\odot }\,{\rm pc^{-2}\,(K\,km\,s^{-1})^{-1}}$ and \alCtwo\ $= 12.5$ $M_{\odot }\,{\rm pc^{-2}\,(K\,km\,s^{-1})^{-1}}$. These values are smaller than our estimated \CI\ conversion factors. 
At this stage, it is difficult to distinguish the different results between each sample are caused by the various estimation methods or by the different intrinsic physical conditions in different galaxies. We hope that more \CI\ observations with higher resolutions (e.g., using ALMA, APEX, \citealt{Krips et al. 2016, Salak et al. 2019, Saito et al. 2020}) will help us to qualify its conversion factor in different galaxy environments.

Using samples of high-redshift gamma-ray burst and quasar molecular gas absorbers with ranges of $z=1.9-3.4$, \citet{Heintz Watson 2020} found that \alCone\ scales linearly with metallicity as: log$\,$\alCone\ $= -1.13\times {\rm log(Z/Z_{\sun})} + 1.33$. They further applied their \alCone\ function for a sample of emission-selected galaxies at $z \sim 0-5$, and found a remarkable agreement between the molecular gas masses inferred from their absorption-derived \alCone\ with the typical \alCO-based estimations. And thus they concluded that the absorption$-$derived \alCone\ can be used to probe the universal properties of molecular gas in the local and high-redshift universe. The simulation in \citet{Glover & Clark 2016} also demonstrated that the \alCone\ scales approximately with metallicity as \alCone\ $ \propto Z^{-1}$ in star-forming clouds. %We have presented the relation of log$\,$\alCone\ $= -1.13\times {\rm log(Z/Z_{\sun})} + 1.33$ as black line in the right-middle panel of Figure\,\ref{fig:acociwithparas}. Our derived \alCone\ distributes around their best-fit linear relation. 

In Figure\,\ref{fig:acociwithheniz2020}, we present our \alCone\ results together with the sample and best-fit linear relation from \citet{Heintz Watson 2020}. 
We further add five star-forming galaxies from ASPECS-LP (short for the ALMA Spectroscopic Survey Large Program; \citealt{Walter et al. 2016, Decarli et al. 2019}) in \citet{Boogaard et al. 2020} for which \Cone\ and CO data is available that have a metallicity in \citet{Boogaard et al. 2019}. The adopted redshift, \LCone, molecular gas mass, and metallicity of each ASPECS galaxy are shown in Table\,\ref{tab:sampleinBooharrd}. \citet{Boogaard et al. 2020} used two methods to estimate  the molecular gas mass separately. The $M_{\rm mol,RJ}$ shown in Table\,\ref{tab:sampleinBooharrd} is estimated via 1.2 mm dust-continuum emission on the Rayleigh-Jeans tail (see the details in their Section 5.4 and Table\,5), and mass of $M_{\rm mol,CO}$ is determined from the CO\,(2$-$1) emission by assuming a luminosity ratio of $L'_{\rm CO(2-1)}/L'_{\rm CO(1-0)} = 0.75 \pm 0.11$ and \alCO$=3.6\,M_{\odot }\,{\rm pc^{-2}\,(K\,km\,s^{-1})^{-1}}$. Using the \LCone\ and 3$\sigma$ for non-detections from Table\,6 in \citet{Boogaard et al. 2020}, we estimate the \alCone\ with both molecular gas masses, and present in Table\,\ref{tab:sampleinBooharrd} and Figure\,\ref{fig:acociwithheniz2020} with colored diamonds.

Though our derived \alCone\ shows almost flat with metallicity, they distribute next to the relation of \citet{Heintz Watson 2020} as seen in Figure\,\ref{fig:acociwithheniz2020}, and similar with the \alCone\ result of ASPECS galaxies as well (the \alCone\ values in both \citealt{Heintz Watson 2020} and \citealt{Boogaard et al. 2020} include a factor of 1.36 to correct the helium and heavier elements).
%our \alCone\ results distribute next to their linear relation, our derived \alCone\ shows almost flat with metallicity with coefficient of $\rho=0.33$.
 The metallicity in our sample mainly focuses in the range of $12 + \rm{log(O/H)} \sim 8.99 - 9.33$, whereas the metallicity of the sample in \citet{Heintz Watson 2020} covers two orders of magnitude in the ranges of $12 + \rm{log(O/H)} \sim 7.12 - 9.15$, which is mostly smaller than our sample. Various studies suggest that \alCO\ increases with decreasing metallicity, turning up sharply below metallicity of $1/3-1/2\,Z_{\sun}$ \citep{Wolfire et al. 2010, Bolatto et al. 2013} where CO is easily photodissociated whereas H$_2$ is self-shields or is shielded by dust from UV photodissociation, and becoming shallower near subsolar metallicity \citep{Glover Mac 2011, Tacconi et al. 2018}. The flat profile between \alCone\ with metallicity in our sample presents that \alCone\ might be similar to \alCO\ and has a fairly shallow metallicity dependence in high metallicity environment. Currently it is difficult to further interpret the correlation between \alCone\ and  metallicity with the small simple size of \alCone. Further studies with higher precision observations spanning a greater range of metallicity might reveal change in \alCone\ with metallicity.
 
Besides, all of the six galaxies are included in the MALATANG (Mapping the dense molecular gas in the strongest star-forming galaxies; Z. Zhang et al. 2021, in prep.) survey. MALATANG is the first systematic survey of the spatially resolved HCN\,(4$-$3) and HCO+\,(4$-$3) emissions in a large sample of nearby galaxies with James Clerk Maxwell Telescope (JCMT). Comparing to both \CI\ emissions, the HCN and HCO+ emissions trace the dense molecular gas that directly relate to SF \citep{Gao Solomon 2004a, Gao Solomon 2004b, Wu et al. 2005, Zhang et al. 2014, Tan et al. 2018, Jiang et al. 2020}. A study of analyzing the carbon excitation and $\alpha_{\rm [CI]}$ with dense molecular gas tracers and SFR using MALATANG survey will be presented in our future works.

\begin{table*}
 % \centering
  \caption{The adopted redshift, molecular gas mass, \Cone\ luminosity, metallicity, and derived \alCone\ for the ASPECS galaxy}
  \label{tab:sampleinBooharrd}
  \begin{threeparttable}
  \begin{tabular}{lcccccccc}
      \hline
     ID & z & $M_{\rm mol,RJ}$ & $M_{\rm mol,CO}$ & $L'_{\rm [CI](1-0)}$ & \alCone$_{\rm ,RJ}$ & \alCone$_{\rm ,CO}$ & Z \\ 
 %   \cline{3-4}\\ 
   \cmidrule(r){3-4} \cmidrule(r){6-7}
     & & \multicolumn{2}{c}{($\times 10^{10} {M}_{\odot}$)} & ($\times 10^{9} {\rm K\,km\,s^{-1}\,pc^2}$) & \multicolumn{2}{c}{($M_{\odot }\,{\rm pc^{-2}\,(K\,km\,s^{-1})^{-1}}$)} \\ 
      \hline
    (1) & (2) & (3) & (4) & (5) & (6) & (7) & (8) \\  
      \hline
1mm.C16 & 1.09 &	 $2.0 \pm 0.2$ & $2.5 \pm 0.4$ & $0.7 \pm 0.2$ & $28.6 \pm 8.7$ & $35.7 \pm 10.8$ & $9.05 \pm 0.08$ \\
3mm.11 & 1.09 &	$\le 0.4$ & $0.6 \pm 0.1$ & $\le 0.3$ & ... & $\ge 20.0$ & $8.78 \pm 0.06$ \\
1mm.C25 & 1.09 & $1.3 \pm 0.2$ & $1.9 \pm 0.5$ & $\le 1.2$ & $\ge 10.8$ & $\ge 15.8$ & $8.70 \pm 0.07$ \\
3mm.16 & 1.29 & $1.2 \pm 0.3$ & $0.9 \pm 0.2$ &	$\le 1.8$ & $\ge 6.7$ & $\ge 5.0$ & >$8.79\pm0.17$ \\
MP.3mm.2 & 1.09	& $\le 0.4$ & $1.1 \pm 0.3$ & $\le 0.9$ & ... & $\ge 12.2$ & $8.73 \pm 0.02$ \\
      \hline

\end{tabular}
\begin{tablenotes}
\footnotesize
\item  Notes: Columns (1)-(5) are adopted from \citet{Boogaard et al. 2020}. (1) and (2) are ASPECS ID and redshift, respectively. Column (3) is molecular gas mass which estimates via the 1.2 mm dust-continuum emission on the Rayleigh-Jeans tail, and (4) is the molecular gas mass determined from the CO\,(2$-$1) emission assuming a luminosity ratio of $L'_{\rm CO(2-1)}/L'_{\rm CO(1-0)} = 0.75 \pm 0.11$ and \alCO$=3.6 M_{\odot }\,{\rm pc^{-2}\,(K\,km\,s^{-1})^{-1}}$ (see Table\,5 in \citealt{Boogaard et al. 2020}). Both of the molecular gas masses from \citet{Boogaard et al. 2020} include a factor of 1.36 to account for heavy elements. Column (5) lists the \LCOone\ with 3$\sigma$ for the non-detections (see Table\,6 in \citealt{Boogaard et al. 2020}). The \alCone\ in columns (6) and (7) are estimated based on molecular gas masses of $M_{\rm mol,RJ}$ and $M_{\rm mol,CO}$, respectively. The metallicity in column (8) is adopted from \citet{Boogaard et al. 2019}.
\end{tablenotes}
    \end{threeparttable}
\end{table*}

\section{Summary and conclusions}

In this paper, we have calibrated the \alCone, \alCtwo, and \alCO\ conversion factors on $\sim1\,$kpc scales for six nearby galaxies using the \CI\ maps observed with $Herschel$ and high resolution maps of \COone, \HI, IR and submm from literatures. We firstly obtained the dust mass using IR and submm data with \citet{Draine Li 2007} dust model. We then adopted three DGR assumptions which all scale approximately with metallicity from precursory results to estimate the gas mass from dust mass. Then combining with \HI\ maps, we are able to solve the conversion factors of \CI\ and CO on $\sim1\,$kpc scales, and estimate the carbon abundance for each system as well. 

We found that similar to the result in \citet{Sandstrom et al. 2013}, \alCO\ shows decreasing in the inner regions of galaxies and becomes flat with galactocentric radii in the galaxy outer regions of galaxies. The central \alCO\ values are on average $\sim 2.2$ times (ranging from $1.1 - 4.2$) lower than the average values of galaxies. The data points for \Cone\ detections are much smaller than those of \Ctwo\ and \COone, and the \alCone\ shows flat with galactocentric radii in our sample. The \alCtwo\ is also mostly flat with galactocentric radii, but the central \alCtwo\ values are slightly lower (ranging from $0.8-2.7$ with mean value of $\sim 1.4$) than the galaxy averages for detections, and become $\sim 1.8$ times (ranging from $0.9 - 3.4$) lower than galaxy averages when considering the limits. The radial profiles of \alCone, \alCtwo, and \alCO\ look similar when using different DGR assumptions. The estimated conversion factors of \alCO, \alCone\ and \alCtwo\ agree with each other well for assumptions of DGR($i$) and DGR($ii$), and both smaller than the values derived with assumption of DGR($iii$). The calibrated carbon abundance shows flat profile with galactocentric radii, and the central and average carbon abundances for each system are comparable. The average carbon abundance of the sample is $X[{\rm CI}]/X[{\rm H_2}] \sim 2.3 \pm 1.1 \times 10^{-5}$, $1.6 \pm 0.7 \times 10^{-5}$, and $1.2 \pm 0.6 \times 10^{-5}$ for the assumption of DGR($i$), DGR($ii$), and DGR($iii$), respectively. And these values are comparable with the widely adopted abundance of $X[{\rm CI}]/X[{\rm H_2}] \sim 3.0 \times 10^{-5}$.

We also presented the CO and \CI\ conversion factors, carbon abundance and excitation temperature $T_\mathrm{ex}$ as functions of the average interstellar radiation field $\overline{U}$, infrared luminosity ${\rm log}(L_{\rm IR})$ obtained from dust model, and metallicity $12 + {\rm log(O/H)}$ from \citet{Moustakas et al. 2010}. We found that \alCO\ has a moderate correlation with ${\rm log}(L_{\rm IR})$ and metallicity, and has no correlation with $\overline{U}$. The \alCone\ shows weak correlation with $\overline{U}$, ${\rm log}(L_{\rm IR})$ and $12 + {\rm log(O/H)}$, while \alCtwo\ only shows week correlation with $\overline{U}$, and has no obvious correlation with other two parameters. We concluded that the \alCone\ might have a fairly shallow metallicity dependence in high metallicity environment, which is similar to \alCO. The carbon abundance weakly correlates with $\overline{U}$, ${\rm log}(L_{\rm IR})$ and $12 + {\rm log(O/H)}$. And the $T_\mathrm{ex}$ has a good, moderate and no correlation with $\overline{U}$, infrared luminosity and metallicity, respectively. We found that among these various correlation analyses that different DGR assumptions give similar results.

\section*{Acknowledgements}

We are grateful to the anonymous referee for an insightful and constructive report.
This work is supported by National Key Basic Research and Development Program of China (grant No. 2017YFA0402704), National Natural Science Foundation of China (NSFC, Nos. 12003070, 12033004, and 11861131007), and Chinese Academy of Sciences Key Research Program of Frontier Sciences (grant No. QYZDJ-SSW-SLH008). Q.J. acknowledges the support by Special research Assistant project. Y.Z. acknowledges the support by the Youth 1000-talent Program of Yunnan Province. %This publication made use of data from COMING, CO Multi-line Imaging of Nearby Galaxies, a legacy project of the Nobeyama 45 m radio telescope.

\section*{Data Availability}

The CO \citep{Kuno et al. 2007} and \HI\ data \citep{Walter et al. 2008} underlying this article can be accessed at websites: {\url{http://www.nro.nao.ac.jp/~nro45mrt/html/COatlas/}} and {\url{http://www.mpia.de/THINGS}}, respectively. The $Spitzer$ and $Herschel$ data is available from NASA/IPAC Infrared Science Archive: {\url{https://irsa.ipac.caltech.edu/frontpage/}}. The derived data generated in this research will be shared on reasonable request to the corresponding author.


%% The best way to enter references is to use BibTeX:
%
%\bibliographystyle{mnras}

%\begin{thebibliography}{99}
%\end{thebibliography}

%%%%%%%%%%%%%%%%%%%%%%%%%%%%%%%%%%%%%%%%%%%%%%%%%%

%%%%%%%%%%%%%%%%% APPENDICES %%%%%%%%%%%%%%%%%%%%%

\appendix
\renewcommand\thefigure{\Alph{section}\arabic{figure}}    

\section{The \CI\ integrated intensity and dust mass distributions for each galaxy}

\begin{figure*}
 \begin{center}
 \includegraphics[bb = 60 147 587 743, width=0.96\textwidth]{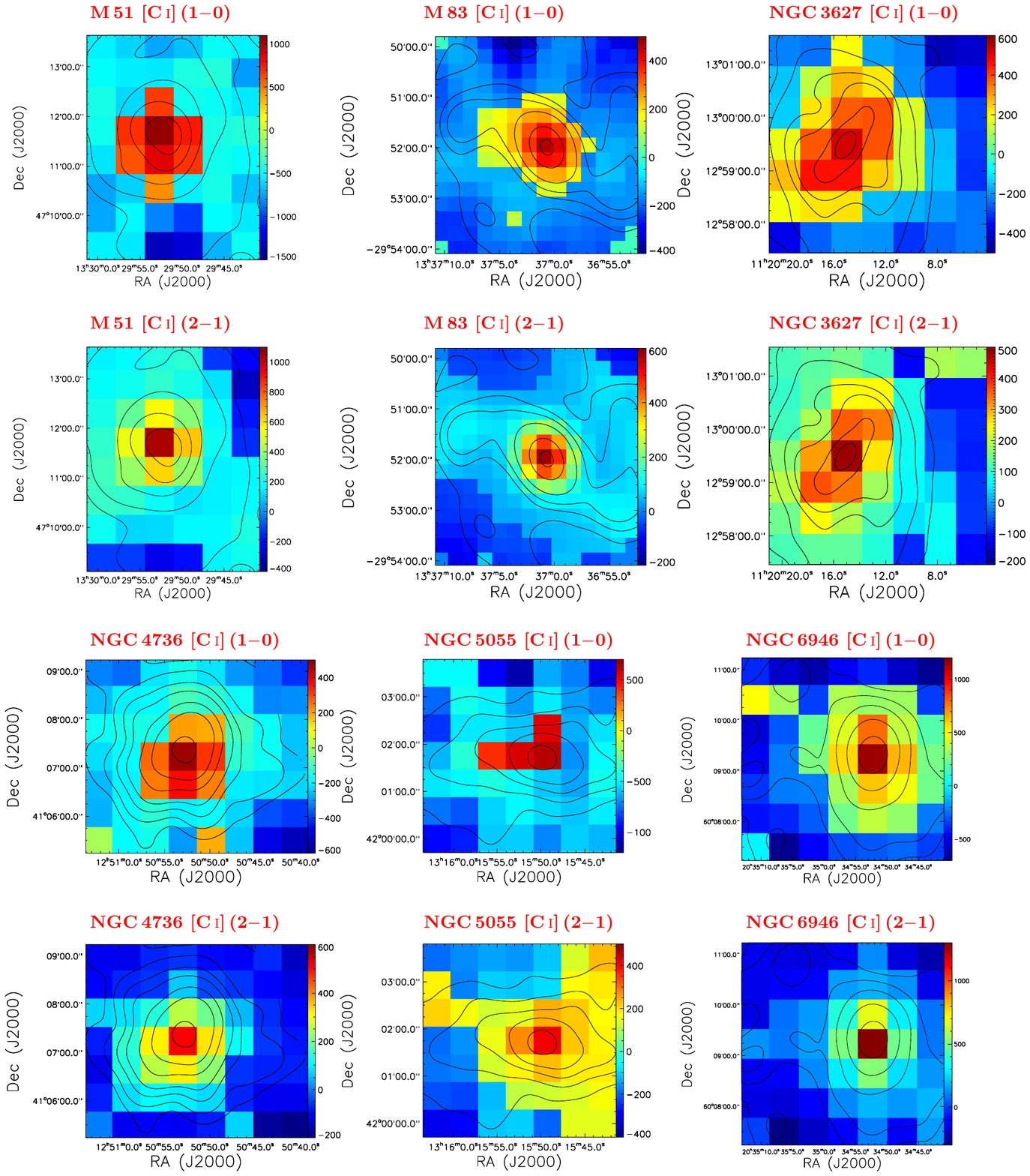}
% {/Users/Qian/Documents/data/Similar_Sandstrom_data/CI_calibration_FTS_layers/CI_emission_All_figure.pdf}
 \end{center}
\caption{The \CI\ integrated intensity (in unit of $\mathrm{Jy\,km\,s^{-1}}$) distribution for each galaxy. The minus values are 3$\sigma$ for the non-detections, and black contours are integrated intensity of \COone\ emissions which have been smoothed to the same resolution as \Cone. The first and third rows are \Cone\ distributions for galaxies of M~51, M~83, NGC~3627, NGC~4736, NGC~5055, and NGC~6946, respectively; the second and forth rows are the corresponding \Ctwo\ distributions for each galaxies. The \COone\ contours are same as \citet{Jiao et al. 2019} for each galaxies: at 10, 40, 100, 200, 300, 400$\sigma$ levels with $\sigma$=0.5 $\mathrm{K\,km\,s^{-1}}$ for M~51; 10, 20, 30, 50, 100, 140$\sigma$ levels with $\sigma$=1.4 $\mathrm{K\,km\,s^{-1}}$ for M~83; 3, 15, 40, 60, 100, 120$\sigma$ levels with $\sigma$=0.9 $\mathrm{K\,km\,s^{-1}}$ for NGC~3627; 3, 30, 100, 200, 400, 600, 1000$\sigma$ levels with $\sigma$=0.05 $\mathrm{K\,km\,s^{-1}}$ for NGC~4736; 5, 10, 15, 20, 25$\sigma$ levels with $\sigma$=2.1 $\mathrm{K\,km\,s^{-1}}$ for NGC~5055; and 8, 16, 24, 40, 80, 160$\sigma$ levels with $\sigma$=1.1 $\mathrm{K\,km\,s^{-1}}$ for NGC~6946, respectively.}\label{fig:CI_All}
\end{figure*}

Figure\,\ref{fig:CI_All} shows the \CI\ integrated intensity distributions with contours of smoothed \COone\ emission adopted from \citet{Kuno et al. 2007}. The minus values are 3$\sigma$ for non-detections. And please find the details of data reduction and distribution in \citet{Jiao et al. 2019}.

\begin{figure*}
 \begin{center}
  \includegraphics[bb = 55 450 580 725, width=0.96\textwidth]{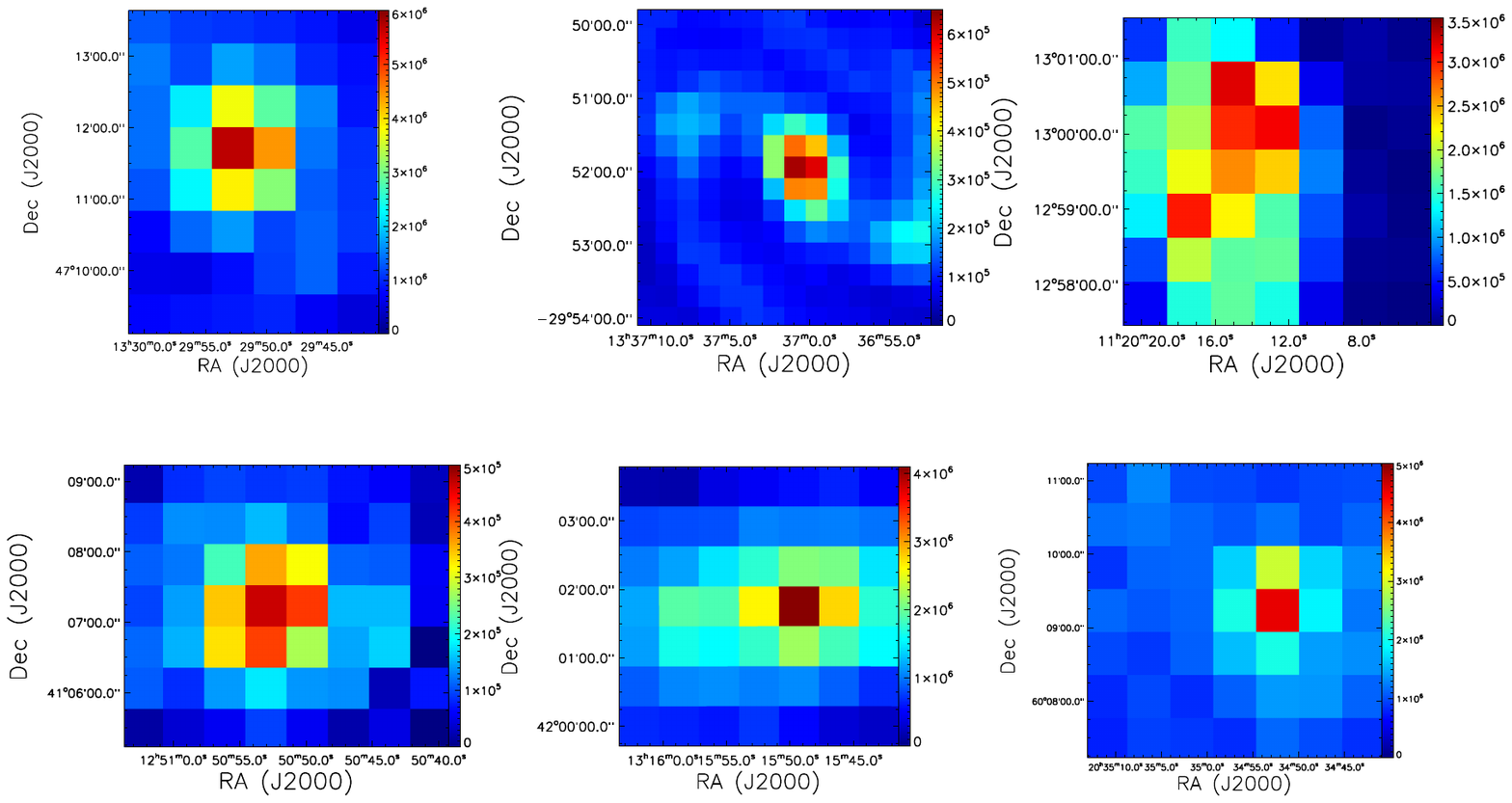}
  %{/Users/Qian/Documents/data/Similar_Sandstrom_data/CI_calibration_FTS_layers/HI_mass_All_figure.pdf}
 \end{center}
\caption{The dust mass distribution in the same region as \CI\ observations. From top left to bottom right panels are galaxies of M~51, M~83, NGC~3627, NGC~4736, NGC~5055, and NGC~6946, respectively.}\label{fig:dustmass}
\end{figure*}

\section{The radii profile of \CI\ and CO conversion factors and carbon abundance with different DGR assumptions}

\begin{figure*}
 \begin{center}
  \includegraphics[bb= 63 190 571 733, width=0.95\textwidth]{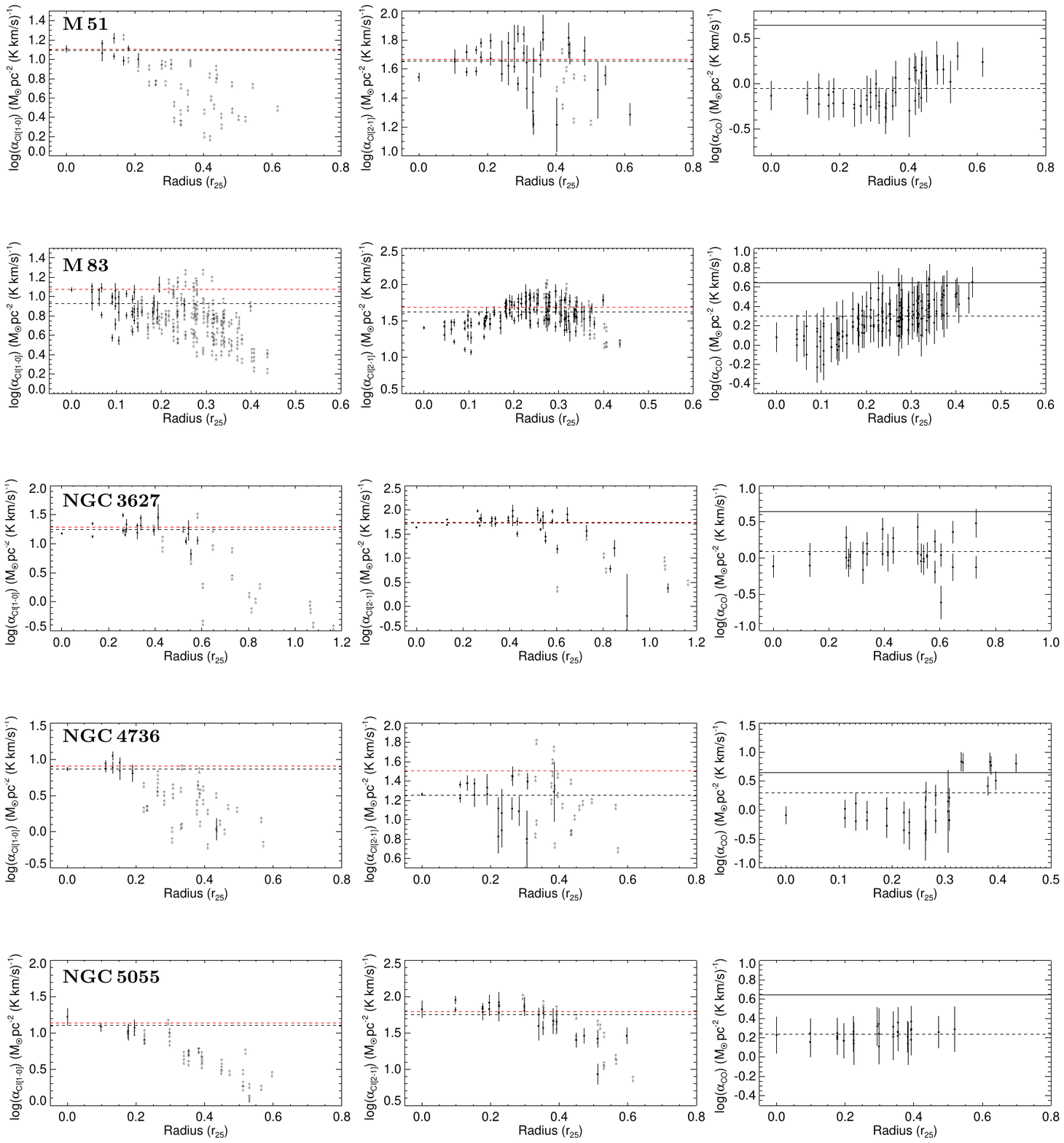} 
 \end{center}
\caption{Same as Figure\ref{fig:aciwithradius} but for M~51, M~83, NGC~3627, NGC~4736, and NGC~5055 from top to bottom, respectively.}
\label{fig:append_radii}
\end{figure*}

\begin{figure*}
 \begin{center}
  \includegraphics[bb= 60 77 569 734, width=0.95\textwidth]{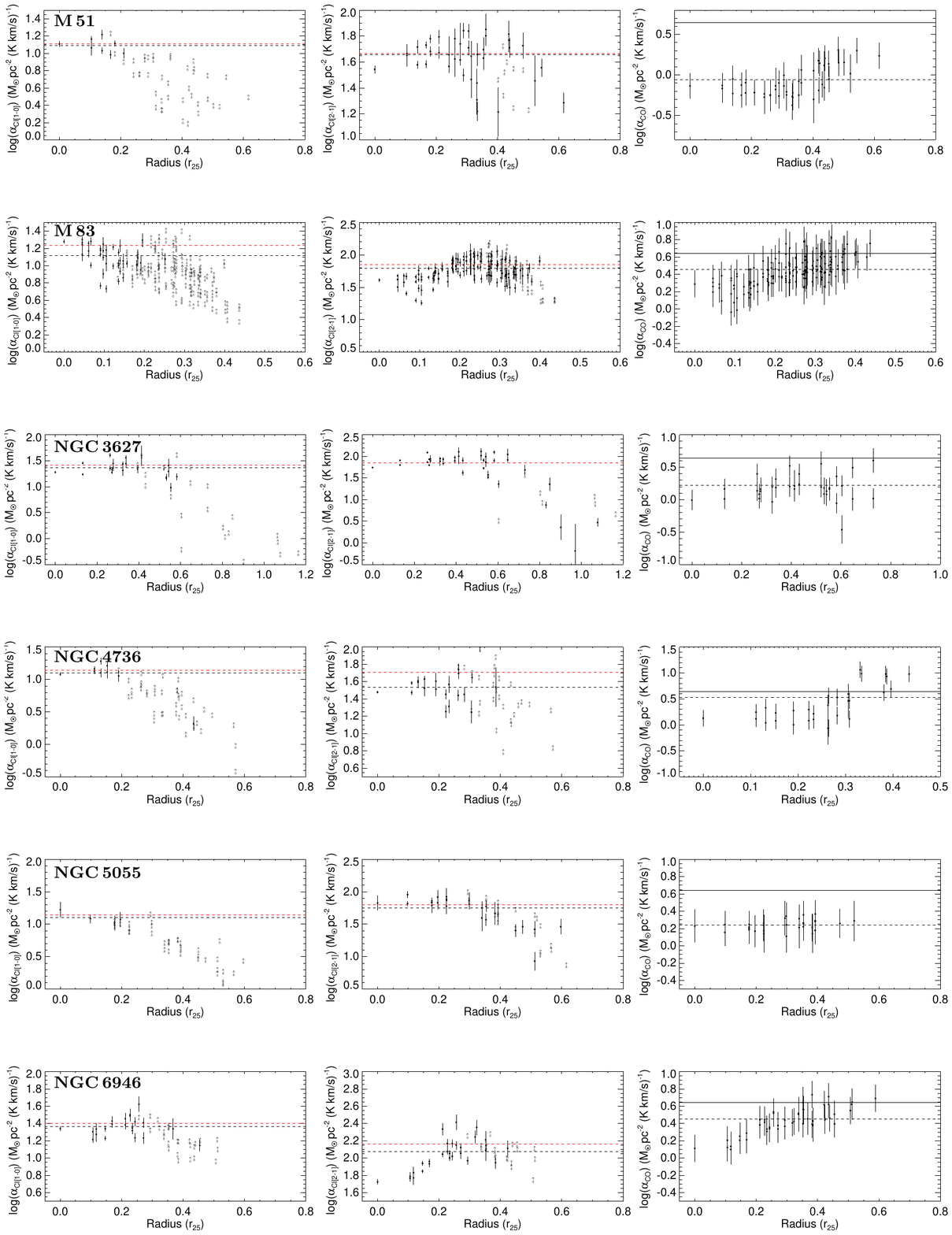} 
 \end{center}
\caption{\alCone\ (left column), \alCtwo\ (middle column), and \alCO\ (right column) as functions of galactocentric radii for M~51, M~83, NGC~3627, NGC~4736, NGC~5055, and NGC~6946 from top to bottom rows with the assumption of DGR$(ii)$, respectively. The symbols are same as in Figure\,\ref{fig:aciwithradius}.}
\label{fig:append_radii_DGRii}
\end{figure*}

\begin{figure*}
 \begin{center}
  \includegraphics[bb= 60 77 569 734, width=0.95\textwidth]{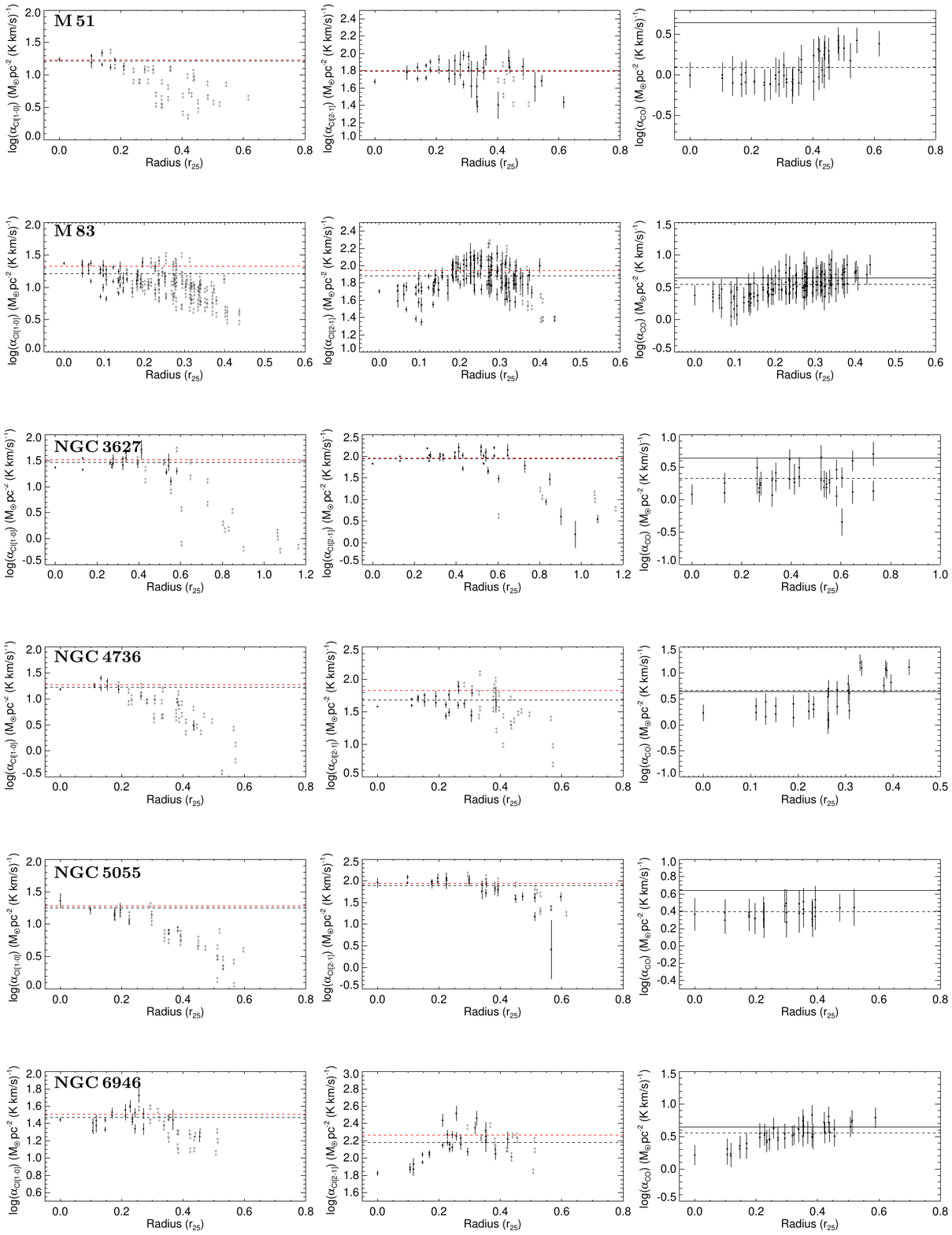} 
 \end{center}
\caption{Same as Figure\ref{fig:append_radii_DGRii} but using the assumption of DGR$(iii)$.}
\label{fig:append_radii_DGRiii}
\end{figure*}

\begin{figure*}
 \begin{center}
  \includegraphics[bb = 55 109 728 487, clip, width=0.47\textwidth]{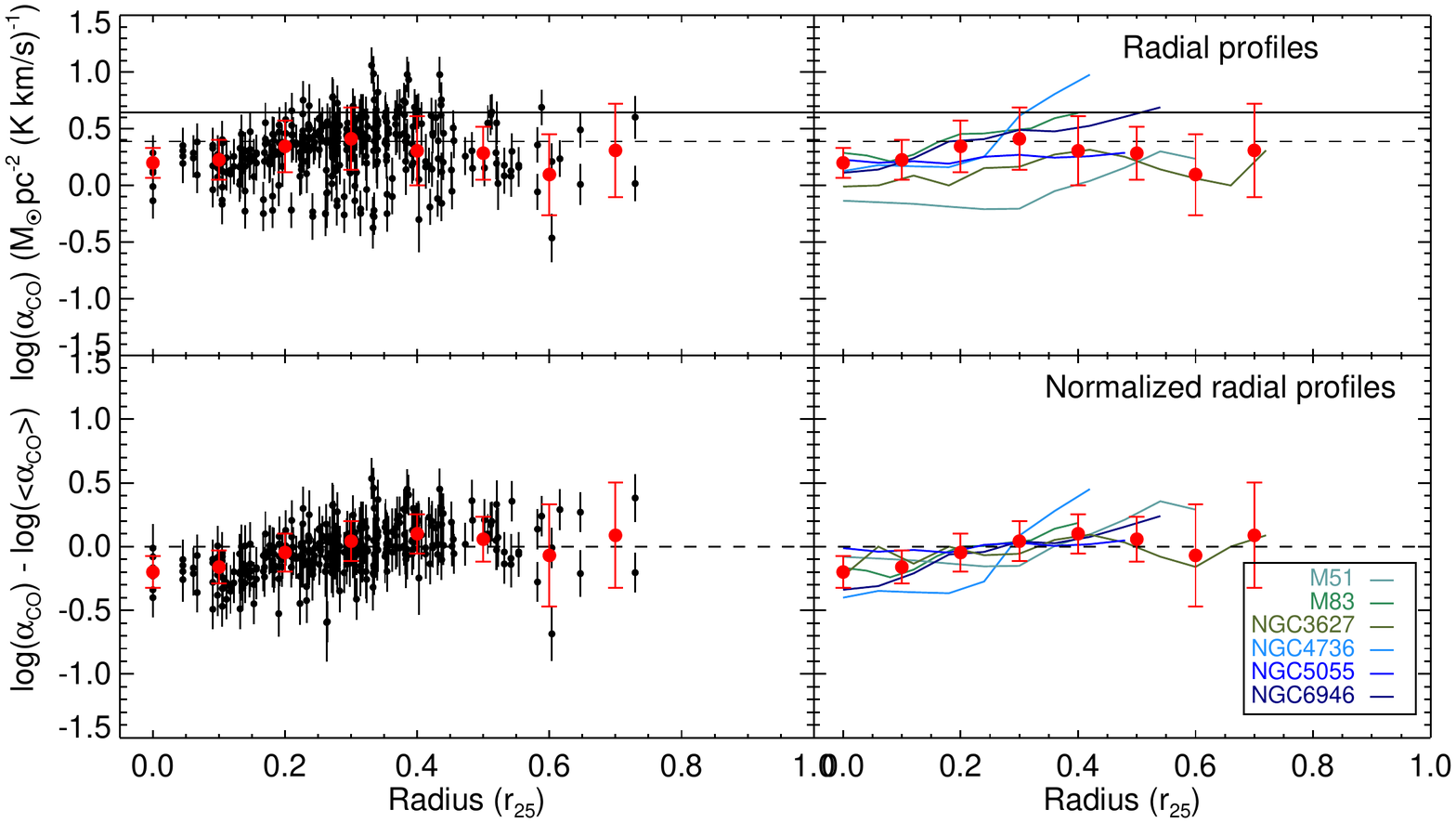} 
  \includegraphics[bb = 55 109 728 487, clip, width=0.47\textwidth]{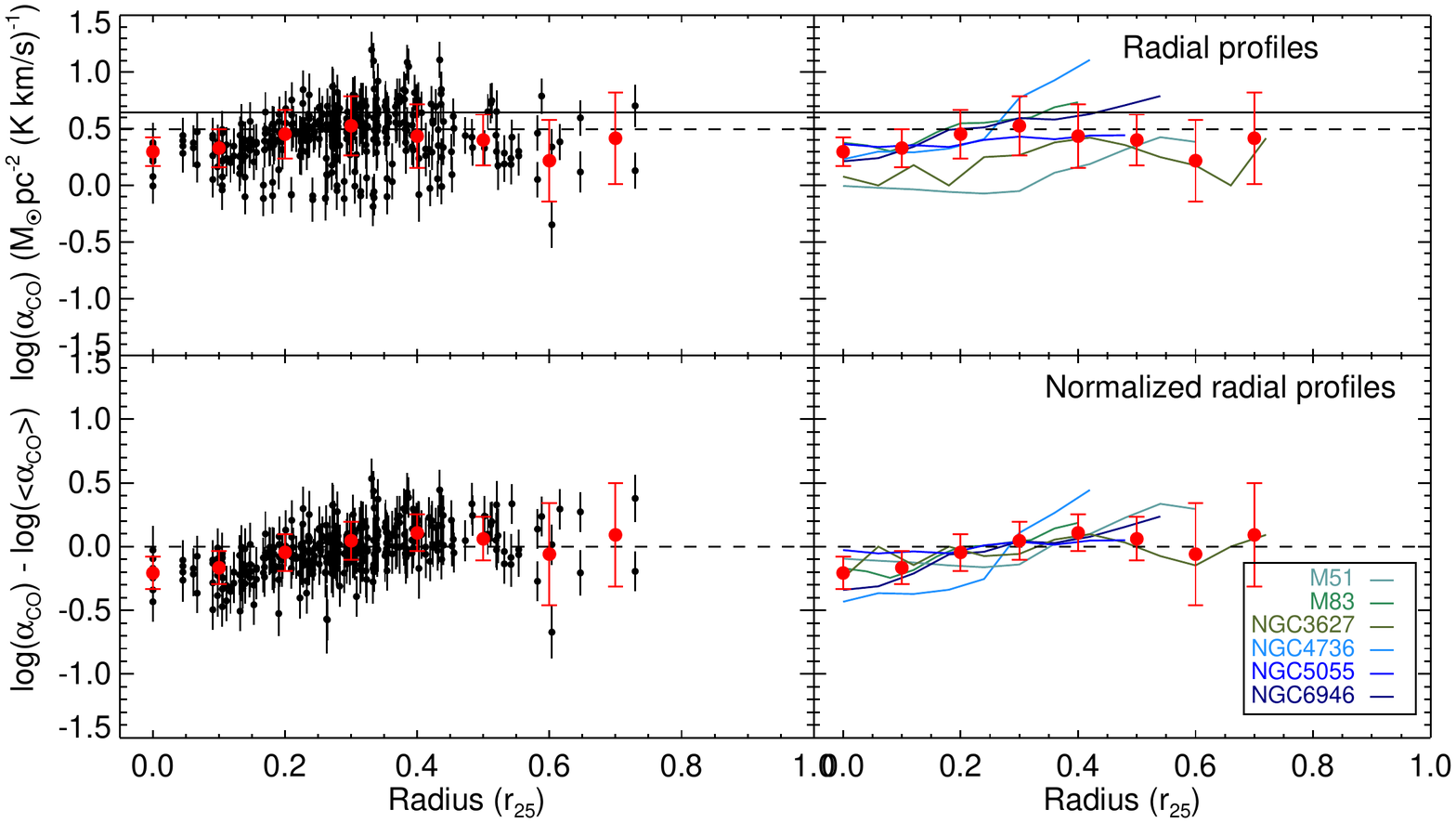}
 \end{center}
\caption{Same as Figure\,\ref{fig:acowithradius_together} but using the assumptions of DGR$(ii)$ (left two columns) and DGR$(iii)$ (right two columns).}
\label{fig:acowithradius_together_DGR}
\end{figure*}

\begin{figure*}
 \begin{center}
  \includegraphics[bb = 61 112 728 486, clip, width=0.47\textwidth]{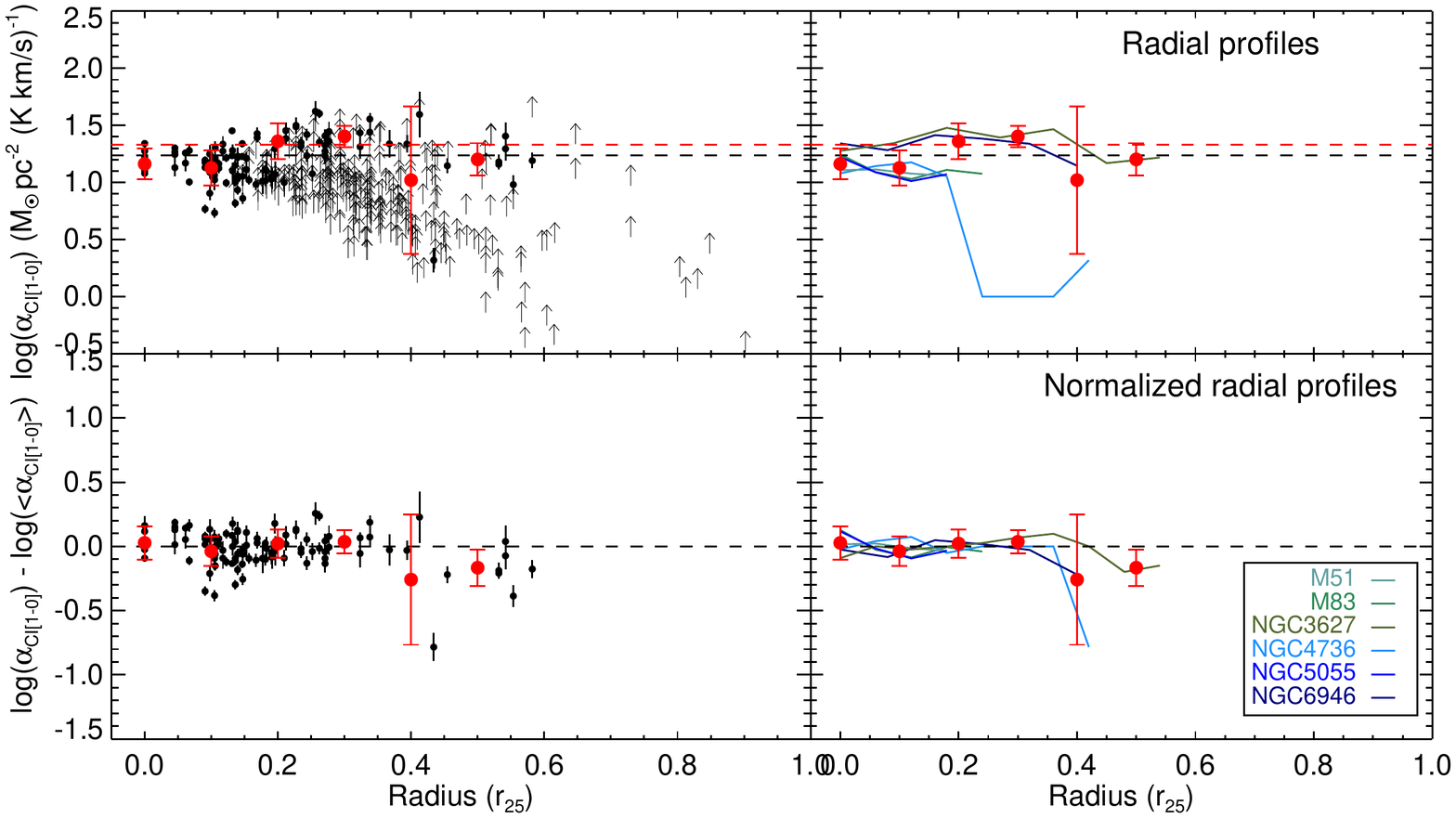} 
  \includegraphics[bb = 61 112 728 486, clip, width=0.47\textwidth]{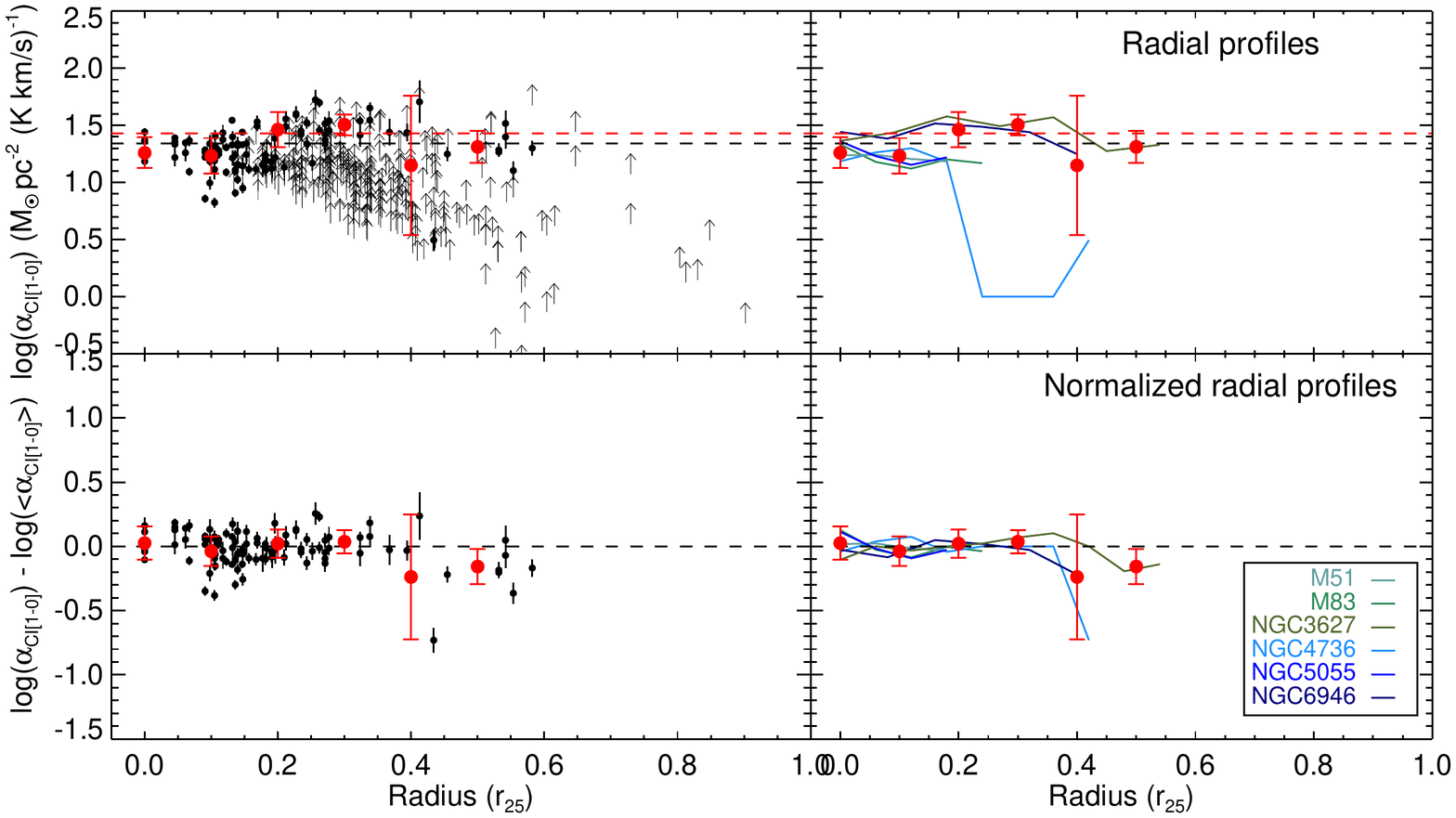} 
 \end{center}
\caption{Same as Figure\,\ref{fig:aciwithradius_together} but using the assumptions of DGR$(ii)$ (left two columns) and DGR$(iii)$ (right two columns).  For the DGR$(ii)$ assumption, the average values are \alCone\ $= 17.3 \pm 7.5 M_{\odot }\,{\rm pc^{-2}\,(K\,km\,s^{-1})^{-1}}$ and \alCone\ $= 21.3 \pm 8.9 M_{\odot }\,{\rm pc^{-2}\,(K\,km\,s^{-1})^{-1}}$ for galaxies together without and with non-detections, respectively. For the DGR$(iii)$ assumption, the average values for galaxies together become \alCone\ $= 21.9 \pm 9.4 M_{\odot }\,{\rm pc^{-2}\,(K\,km\,s^{-1})^{-1}}$ and \alCone\ $= 27.0 \pm 11.3 M_{\odot }\,{\rm pc^{-2}\,(K\,km\,s^{-1})^{-1}}$ without and with non-detections, respectively.}
\label{fig:aciwithradius_together_DGR}
\end{figure*}

\begin{figure*}
 \begin{center}
  \includegraphics[bb = 61 112 728 486, clip, width=0.47\textwidth]{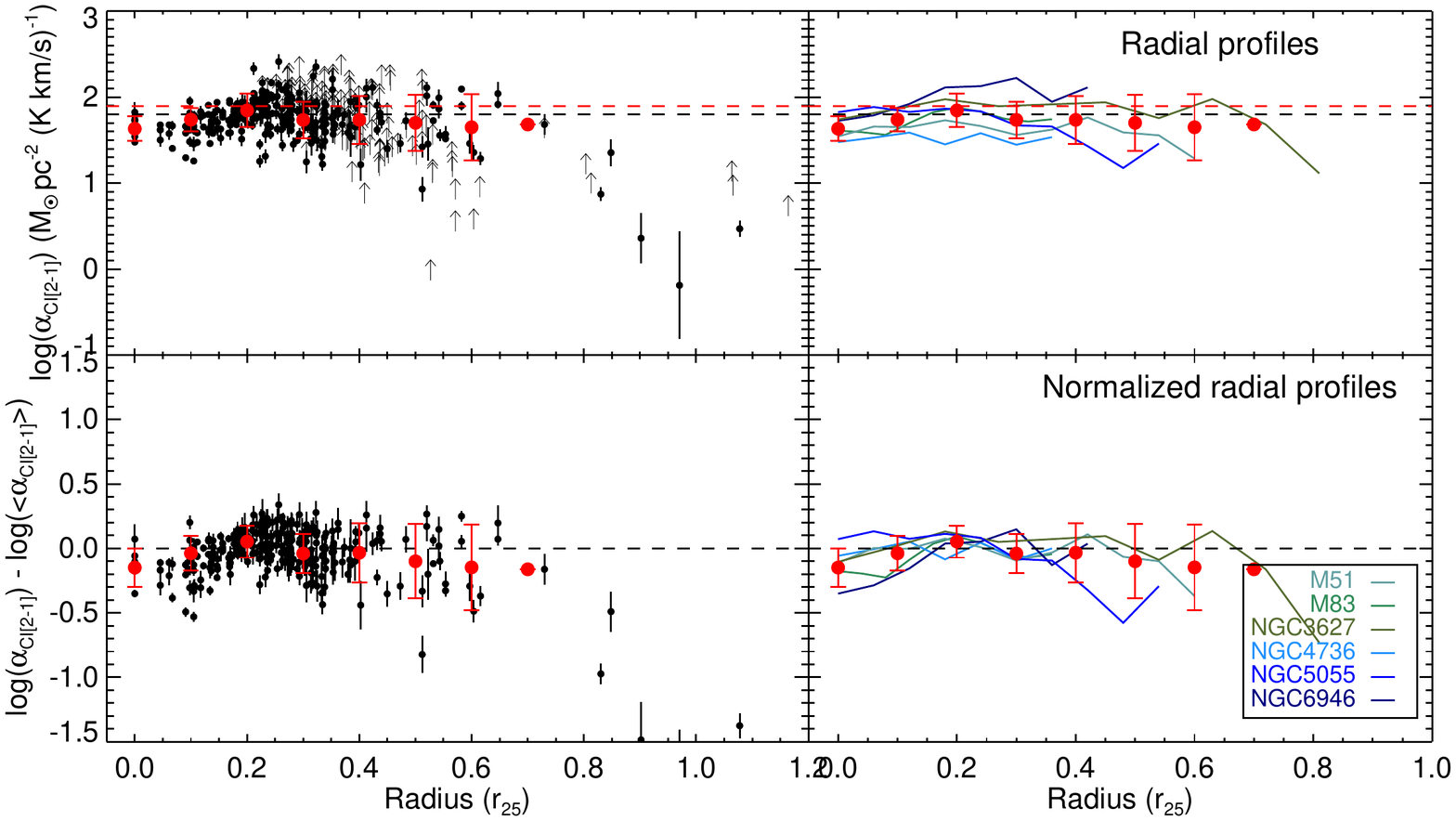} 
  \includegraphics[bb = 61 112 728 486, clip, width=0.47\textwidth]{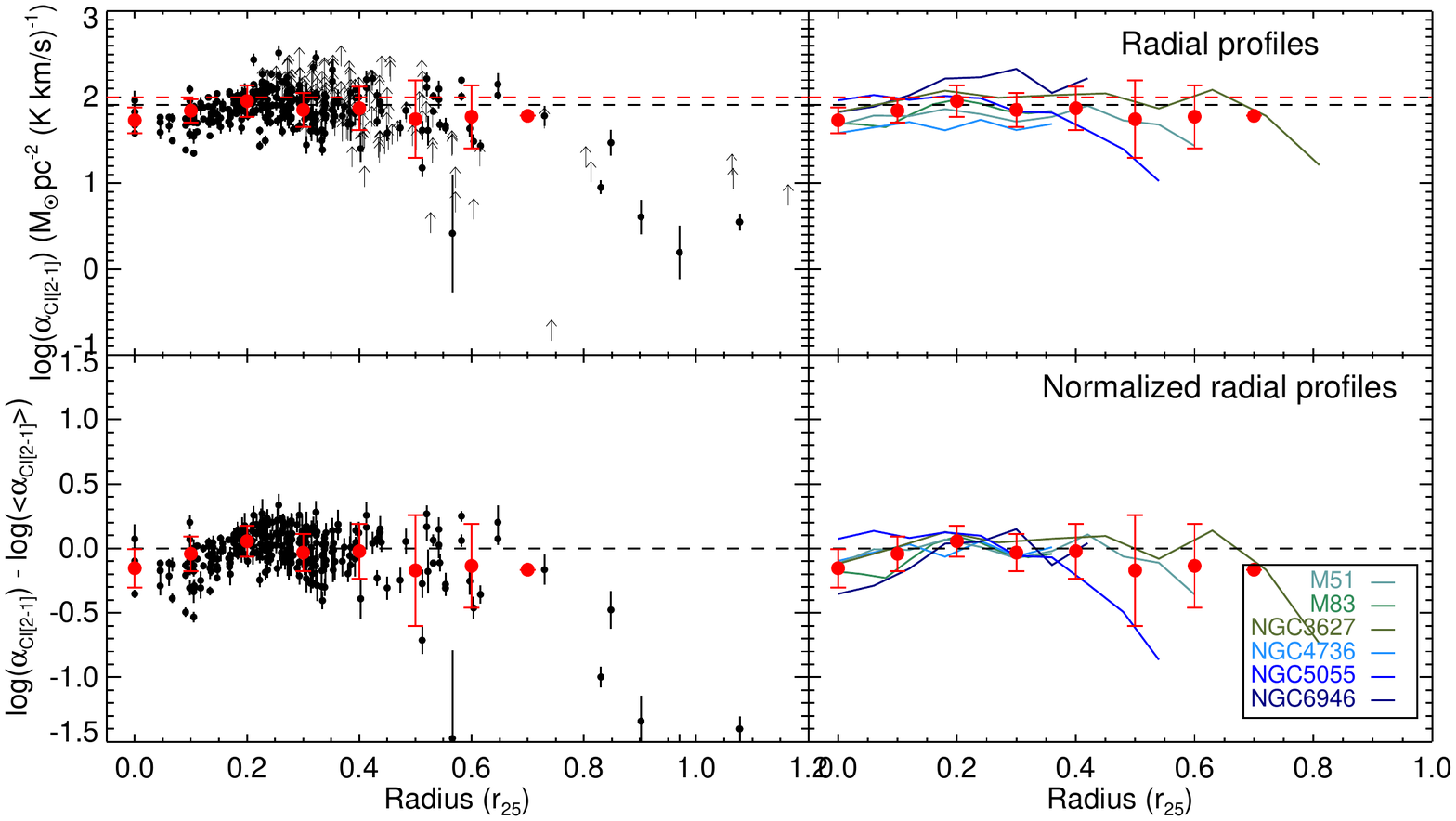}
 \end{center}
\caption{Same as Figure\,\ref{fig:aciwithradius_together_DGR} but for \alCtwo. For the DGR$(ii)$ assumption, the average values are \alCtwo\ $= 63.6 \pm 32.9 M_{\odot }\,{\rm pc^{-2}\,(K\,km\,s^{-1})^{-1}}$ and \alCtwo\ $= 78.8 \pm 49.1 M_{\odot }\,{\rm pc^{-2}\,(K\,km\,s^{-1})^{-1}}$ for galaxies together without and with non-detections, respectively. For the DGR$(iii)$ assumption, the average values for galaxies together become \alCtwo\ $= 81.2 \pm 41.3 M_{\odot }\,{\rm pc^{-2}\,(K\,km\,s^{-1})^{-1}}$ and \alCtwo\ $= 100.5 \pm 61.8 M_{\odot }\,{\rm pc^{-2}\,(K\,km\,s^{-1})^{-1}}$ without and with non-detections, respectively.}
\label{fig:aci21withradius_together_DGR}
\end{figure*}

\begin{figure*}
 \begin{center}
  \includegraphics[bb = 62 110 727 486, clip, width=0.47\textwidth]{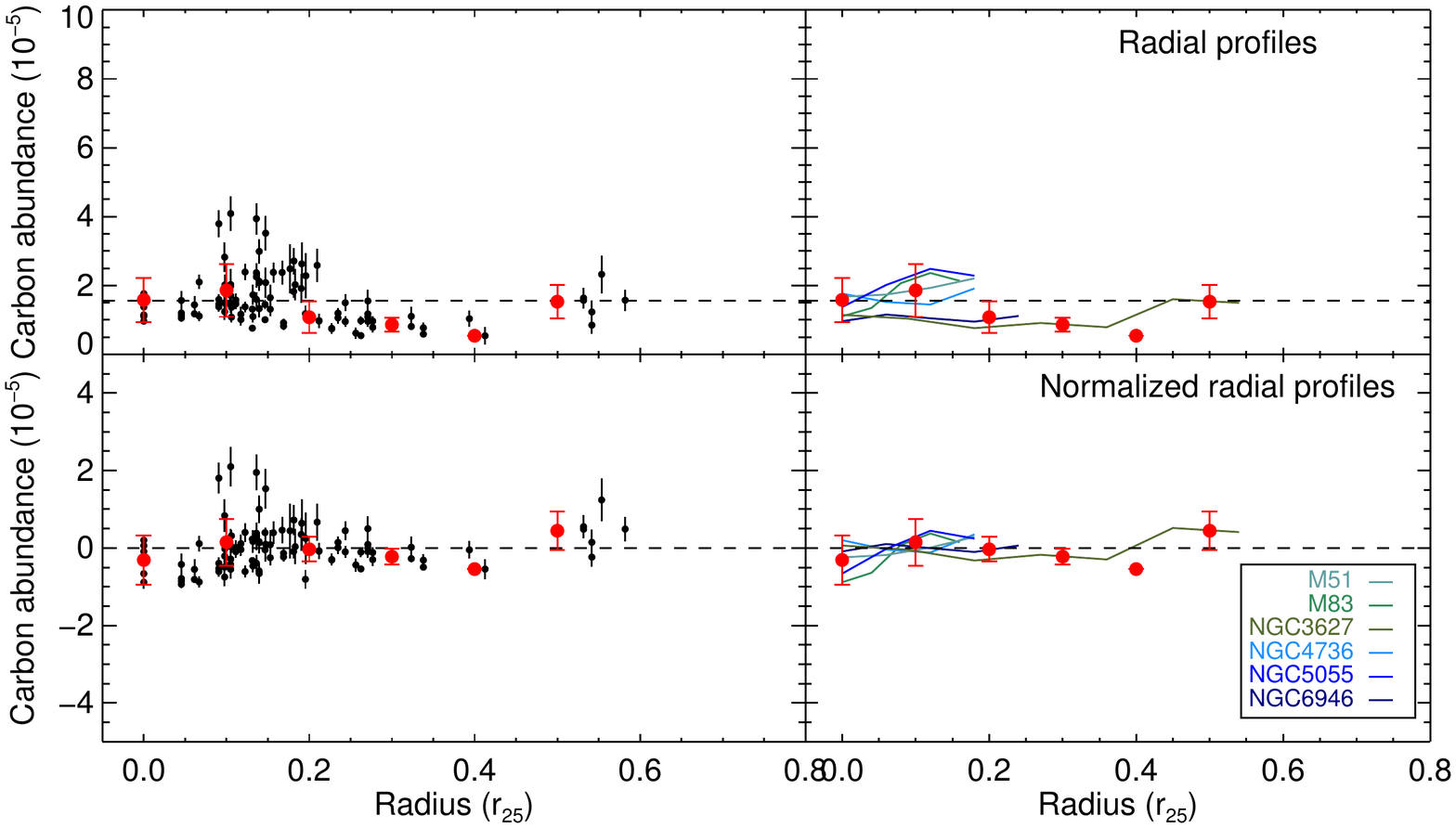} 
  \includegraphics[bb = 62 110 727 486, clip, width=0.47\textwidth]{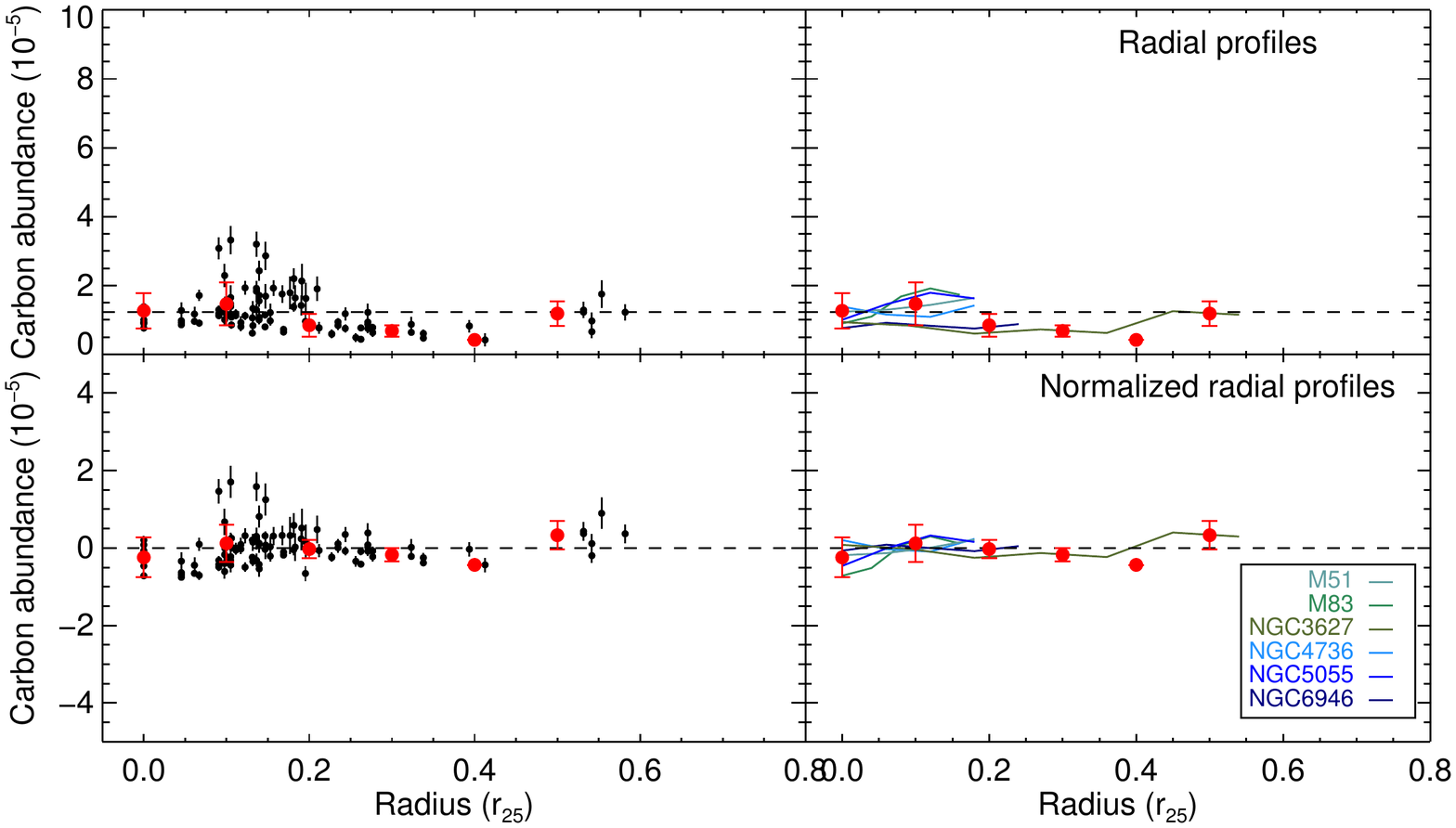}
 \end{center}
\caption{Same as Figure\,\ref{fig:acowithradius_together_DGR} but for carbon abundance.}
\label{fig:ciabundancewithradius_together_DGR}
\end{figure*}

%%%%%%%%%%%%%%%%%%%%%%%%%%%%%%%%%%%%%%%%%%%%%%%%%%

% Don't change these lines
\bsp	% typesetting comment
\label{lastpage}
\end{document}